\documentstyle[aps,epsfig,floats,amssymb]{revtex}
\draft

\newcommand{\lsi}{\raise0.3ex\hbox{$<$\kern-0.75em\raise-1.1ex\hbox{$\sim$}}}
\newcommand{\gsi}{\raise0.3ex\hbox{$>$\kern-0.75em\raise-1.1ex\hbox{$\sim$}}}
\begin{document}
\twocolumn[\hsize\textwidth\columnwidth\hsize\csname
@twocolumnfalse\endcsname

\title{Probing Quintessence with Time Variation of Couplings}

\author{Christof Wetterich}

\address{
Institut f{\"u}r Theoretische Physik,
Philosophenweg 16, 69120 Heidelberg, Germany}

\maketitle

\begin{abstract}
Many models of quintessence predict a time variation of the fundamental
constants as well as a composition-dependent gravity like
long-range force mediated by the cosmon. We present bounds for
the cosmon coupling to matter and radiation within a grand unified framework.
The unification scale, the unified gauge coupling and the Fermi scale are allowed to
vary independently. We find that the variation of the weak scale compared to the nucleon
mass is
severely restricted. The violation of the equivalence principle turns out to be substantially
larger than in models where only the electromagnetic fine structure constant varies with
time. We also show that in contrast to gravity the local cosmon field in a
condensed object does not decouple from the cosmological evolution. In consequence, the
cosmon interaction constitutes a possible quantitative
link between cosmological observations and several areas of
high precision experiments concerning the local time or space variation of couplings and
tests of the equivalence principle.
\end{abstract}
\pacs{PACS numbers: 98.80.Cq,04.50.+h,95.35.+d,04.80.-y  \hfill HD-THEP-02-11}

 ]


\section{Introduction}
Over the last years several observations point to a universe which
is dominated by some form of homogeneously distributed dark energy.
An interesting candidate for dark energy is quintessence -- the energy
density of a slowly evolving scalar field \cite{CW2,PR,Q}.
It is a characteristic feature of the quintessence scenario that fundamental
coupling constants depend on time even in late cosmology
where such a time variation could be observable \cite{CW2,SF,CW3,DZ,CK}.
This effect is quite generic: the couplings
depend on the expectation value of the scalar ``cosmon'' field, which
in turn varies in the course of its cosmological evolution. The size of this
effect depends, however, on physics at the unification scale \cite{CW2} and
is not known at present. An observation of time-dependent
fundamental ``constants'' could be interpreted as a signal in favor of
quintessence -- no such time dependence would be connected to dark energy if the latter
occurs in the form of a cosmological constant.

Recently, a low value of the electromagnetic fine structure constant
has been reported \cite{Obs} for absorption lines in the light from
distant quasars. The data are consistent with a variation $\Delta
\alpha_{em}/\alpha_{em}\approx-0.7\cdot 10^{-5}$ for a redshift
$z\approx 2$. This has triggered renewed interest in the theoretical
issues related to time varying fundamental couplings  \cite{DZ,CK,AV,SBM,OP,CFR,LSS,DF,Uz},
a subject that has been pioneered long ago \cite{DMJ,Jo} and explored in various
contexts \cite{LL}. Concerning the observational evidence, a definite conclusion
seems premature to us -- the discussion of this paper can easily
be adapted to a smaller time
variation of the fundamental constants as well. Nevertheless, taking
the reported time variation of $\alpha_{em}$ at face value would
fix the coupling strength of the cosmon to matter and radiation.

Typical models of quintessence relate the time variation of $\alpha_{em}$
to the time variation of other fundamental parameters,
composition-dependent gravity-like forces, a dependence of couplings
on the distance from a massive body like the earth and cosmological parameters
of dark energy. These links arise on three layers. Logically, the time variation of
couplings needs not to be related to the time variation of a scalar field. On this first
layer one can nevertheless find connections between the variations of dimensionless
couplings like the fine structure constant $\alpha_{em}$ and mass ratios like the nucleon
to Planck mass. The present work is based on the hypothesis of grand unification. A second
layer assumes that the time variation of couplings originates from a scalar field with mass
much smaller than the inverse extension of our solar system. On this level we encounter a
connection with tests of the equivalence principle. The quantitative relation between the
variation of couplings and the violation of the equivalence principle depends, however,
on the rate of cosmological change of the scalar field. In general, the scalar field needs
not to be related to dark energy. For example, coupling variations due to a ``runaway
dilaton'' in string theories \cite{DV} have not necessarily a sizeable impact on dark
energy. If the contribution of the scalar field to the energy density of the universe
is very small the rate of change in time is not directly accessible to cosmological
observation. The third layer finally postulates that the relevant scalar field is the
cosmon whose kinetic and potential energy constitute the dark energy of the universe.
The differential acceleration of two test bodies with equal mass can now be
quantitatively related to the time variation of $\alpha_{em}$ in terms of observable
cosmological parameters.

In a quintessence scenario the variation of the gauge
couplings can arise from a coupling of the cosmon field
$\chi$ to the kinetic term for the gauge fields in a grand unified theory \cite{HSW}
\begin{equation}\label{1.1}
{\cal L}_F=\frac{1}{4}Z_F(\chi)F^{\mu\nu}F_{\mu\nu}\end{equation}
Such a coupling preserves all symmetries \footnote{See \cite{Jo,Bek} for gauge invariant
formulations of a time dependent fine structure constant in the context of QED.}
and makes the renormalized
gauge coupling $g\sim Z_F^{-1/2}$ dependent on $\chi$, and therefore
on time if $\chi$ evolves. In a grand unified theory the time
variation of the electromagnetic fine structure constant is directly
related to a time variation of the strong gauge coupling and therefore
also to a time variation of the nucleon mass $m_n$ \cite{EOW,CFR,LSS,DF}.
We discuss the time variation of $\alpha_{em}$ and $m_n$
in the case where both the unified coupling $g(M_{GUT})$ and the ratio between the
unification scale $M_{GUT}$ and the Planck mass $\bar{M}_p$ may depend on time.
Furthermore, one expects a time variation of fermion masses and the weak
interaction scale. We include here a possible time variation of the Fermi scale or,
more precisely, $M_W/M_{GUT}$, keeping the Yukawa couplings fixed for simplicity.
This influences the relation between the time variation of $\alpha_{em}$ and
$m_n/\bar{M}_p$. As a result we present bounds on the time variation of $g(M_{GUT}),
M_{GUT}/\bar{M}_p$ and $M_W/M_{GUT}$. We are not aware of a previous investigation
taking all these effects simultaneously into account.

The cosmon field depends both on time and space coordinates. For a
massive body like the earth $\chi$ depends on the distance from the
center. Therefore also the fundamental couplings will depend on the
location of the probe \cite{EOW}. Their values in space differ from those measured
at the earth surface. Furthermore, a  spacially varying scalar field
coupling to matter mediates a force. Since the cosmon is effectively massless on the
scale of our solar system, this force resembles gravity in many
respects \cite{CW2}, \cite{CW3}. However, if the cosmon coupling
to matter is not precisely proportional to mass, the new gravity-like
force will depend on the composition of the test bodies and therefore
appear as a violation of the equivalence principle \cite{PSW}. The
existing severe bounds on such violations of the equivalence
principle \cite{EOT} imply strong restrictions on the size
of the cosmon coupling to matter \cite{CW3,DZ,CK}, which we take
into account for our bounds on the time variation of couplings. In particular,
we find that the violations of the equivalence principle are substantially larger in
a grand unified setting as compared to estimates \cite{DZ,CK} where only the
effect of the variation of the fine structure constant is taken into account.

Since the space and time variation of a scalar field originate from the same kinetic
term in the effective action we can express the differential acceleration $\eta$ of two
test bodies with equal mass in terms of the dependence of $\alpha_{em}$ on redshift
$z$
\begin{equation}\label{1BB}
\eta=-1.75\cdot 10^{-2}\left(\frac{\partial\ln\alpha_{em}}{\partial z}\right)^2_{|z=0}
\frac{\Delta R_Z(1+\tilde{Q})}{\Omega^{(0)}_h(1+w^{(0)}_h)}.
\end{equation}
Here $\Delta R_Z=\Delta Z/(Z+N)\approx0.1$ for typical experimental tests of the
equivalence principle. The quantity $\tilde{Q}$ accounts for the influence of the
variation of mass ratios like $M_W/m_n,m_e/m_n,M_{GUT}/\bar{M}$ beyond the variation of
$m_n/\bar{M}$ which is related to the variation of $\alpha_{em}$ by grand unification.
With $\partial\ln\alpha_{em}/\partial\ln z\approx 10^{-6}$ saturating the bound from the
Oklo natural reactor, $\eta$ comes close to the present detection limit if the present
equation of state of quintessence $w^{(0)}_h$ is close to $-1$ as suggested by
observation. The effect is enhanced by a large value of $\tilde{Q}$ unless the variation
of $M_W/m_n$ is severely restricted. We also note that the present energy density of the
scalar field $\Omega^{(0)}_h$ cannot be very small if the QSO-observation of a varying fine
structure constant is required to be compatible with the present bound $\eta<3\cdot 10^{-13}$
\cite{EOT}. Discarding accidental cancellations the QSO observation seems to point towards an
association of the scalar field with the cosmon.

Finally, if masses and couplings depend on time, this could also influence
the cosmological evolution. For example, this effect modifies
the conservation of the energy momentum tensor of matter \cite{SF,CW2,CW3}.
In view of the bounds of this paper this effect is very small for ordinary matter
\cite{CW3} - a larger coupling to dark matter remains possible, however.

A time variation of couplings may also be observed by local experiments on earth
\cite{Oklo} or within our solar system. This raises the important question to what
extent the local time variation is related to the cosmological evolution of the
cosmon field $\chi$. As far as gravity is concerned a bound local system decouples
from the cosmological evolution - we cannot measure the time evolution of the
cosmological scale factor by terrestial experiments. We show in this paper (sect. 9)
that the situation is very different for the evolution of the cosmon field. Indeed, the
values of the fundamental couplings measured on earth or in the solar system depend
directly on the cosmological value of $\chi$. For the time evolution of couplings the time
dependence of $\chi(t)$ acts as a universal clock! This important difference with respect
to gravity can be traced back to the presence of an effective potential for the cosmon.

In this paper we relate all the different aspects of the cosmon coupling
to matter and radiation quantitatively. We derive bounds on the variation of
$M_{GUT}/\bar{M}_p$ and $M_W/m_n$ under the hypothesis that the QSO observation of a
varying $\alpha_{em}$ fixes the size of the cosmon coupling to radiation and matter
and therefore the variation of the unified gauge coupling and $m_n/\bar{M}_p$. We
investigate the effects of a separate variation of the grand unified gauge coupling and the
scale of electroweak symmetry breaking or, equivalently, of the
characteristic scales of strong and weak interactions. We also account
for a possible variation in the ratio between the Planck mass and the
scale of grand unification. In this generality our discussion goes beyond
previous phenomenological work \cite{DZ,CK,CFR,LSS,DF}. For simplicity of the presentation we have
associated all fermion masses and the mass of the $W$-boson to a common
weak interaction scale. This should only serve as a guide -- the dependence
of these masses on the cosmon field is expected to be more complicated.
The corresponding bounds should therefore be used with caution.

An overall picture emerges where quintessence can reconcile
the reported time variation of $\alpha_{em}$ from quasar absorption lines with the
present bounds from other observations. However this holds
only for a particular class of models - the
cosmic evolution of the scalar field must have slowed down considerably between
$z\approx 2$ and $z<0.5$.
For example, combining the QSO value of $\Delta\alpha_{em}$ with
a corresponding bound from the Oklo natural reactor leads to a bound \footnote{
See sect. 9 for cautious remarks concerning the validity of the information
about $\Delta\alpha_{em}$. This bound holds also for the equation of state of a scalar
field which is responsible for the time variation of couplings without being relevant
for dark energy} for the present equation of state of quintessence
\begin{equation}\label{AA2}
w_h=p_h/\rho_h<-0.9
\end{equation}
This bound is comparable with the most severe bounds from cosmological observations.
Typically, the order of magnitude of the cosmon coupling
turns out such that its effects may soon be seen by other experiments
-- or the reported time variation of $\alpha_{em}$ may be excluded,
at least within the quintessence scenario.

Already now the small size of the cosmon coupling to matter demands for a natural
explanation. (This issue would get even more severe if future observations exclude a
time variation of couplings with the presently inferred strength.)
A recent proposal in this direction
invokes that the couplings may be closed to a fixed point of their evolution equations
\cite{CWCF}. It is the aim of this paper to draw the attention on the fact that generic
models of quintessence based on the late time evolution of a scalar field have to be
confronted with the possibility of time varying fundamental constants. There are by
now many proposals for obtaining quintessence from some more fundamental considerations.
Very few, however, discuss the problem of time varying couplings. The bounds to be
respected are very tight and severely constrain the allowed couplings of the scalar field
to other forms of matter and radiation. A time variation of the cosmon field necesarily
occurs whenever the equation of state for quintessence $w_h$
differs from $-1$. (For $w_h=-1$
one recovers the phenomenology of a cosmological constant.) For specific proposals of
quintessence within string theories or similar unified theories the overall size of the
cosmon couplings may not be a free parameter. For $w_h\neq -1$ a careful discussion of the
issue of time varying couplings becomes then mandatory.

The computation of the time variation of couplings in a given model of quintessence involves
two steps. First, one needs the dependence of dimensionless couplings and mass ratios on the
cosmon field $\chi$. At this point the specific proposal for the role of the cosmon field
within a unified theory enters. We explore here the simple assumption that the couplings
depend on the cosmon field $\chi$ logarithmically. In this case we can
parameterize the most important
quantities relevant for our problem by three parameters. The second step needs the time
dependence of the cosmon field, thereby translating the field dependence of couplings into a
time dependence. Here the cosmological equations for the cosmon - in particular, the shape
of the cosmon potential and kinetic term - play a determinant role. For the latest
cosmological epoch this can be parameterized by an equation of state for quintessence.

Our paper is organized as follows.
In sects. 2 and 3 we discuss the field dependence of couplings and
mass ratios within the setting of a grand unified theory. There we
use a language where all mass scales depend on the cosmon field.
This is translated to the more familiar Einstein frame with fixed Planck
mass in sect. 4. In this section we also discuss the cosmological time
dependence of the cosmon field. For this purpose we investigate a very general class
of quintessence models where the scalar kinetic term contains two derivatives and the
potential vanishes monotonically for large fields.

In the following sections we address the
different facets how the cosmon interactions may be observed.
In sect. 5 we compute the strength of the cosmon
coupling to matter and radiation from the reported time variation of the
fine structure constant. Sect. 6 discusses consequences for nucleosynthesis
and in sect. 7 we address the tests of the equivalence principle. In sect.
8 we collect the bounds on the field dependence of gauge couplings and
mass scales. Sect. 9 turns to terrestial and satellite
observations of a time or space
dependence of fundamental couplings. We show that the time dependence
is universal -- it should be the same on earth (or other objects
decoupled from the gravitational expansion) and for cosmological
observations. Our conclusions are presented in sect. 10.

\section{Field-dependent couplings}
For a discussion of the origin of the field
dependence of couplings we employ a language which is
convenient for an understanding of dynamical mass scales.
All mass scales including the Planck mass depend on the cosmon
field $\chi$. This language highlights the basic observation that physical
observables can only depend on dimensionless couplings and ratios of mass
scales. We will later (sect. 4) switch to the more familiar language where
the Planck mass is kept fixed.
We use an (euclidean) action
\begin{equation}\label{A}
S=\int d^4x\sqrt{\bar g}({\cal{L}}_q+{\cal{L}}_m)\end{equation}
where the gravity and quintessence part
\begin{equation}\label{B}
{\cal{L}}_q=-\frac{1}{12}f^2(\chi)\chi^2\bar R+\frac{1}{2} Z_\chi(\chi)
\partial_\mu\chi\partial_\nu\chi\bar g^{\mu\nu}+V(\chi)
\end{equation}
describes the dynamics of the metric
$\bar g_{\mu\nu}$ and the scalar cosmon field $\chi$,
whereas ${\cal{L}}_m$ accounts for the ``matter'' and its couplings.
Typically, the value of $\chi$ will change during the evolution of the universe
and therefore the coefficient in front of the curvature scalar or
the effective Planck mass
\begin{equation}\label{2.2a}
M^2_p(\chi)=\frac{4\pi}{3}f^2\chi^2\end{equation}
depends on time in this formulation.
The field $\chi$ is associated to a dynamical unification scale,
as, for example, in string theories.

For the matter part, let us
discuss for definiteness a grand unified theory based on $SU(5)$
or $SO(10)$. Then ${\cal{L}}_m$ contains a covariant kinetic term for the
gauge fields $A^z_\mu$ and fermions $\psi$
\begin{equation}\label{C}
{\cal{L}}_m=\frac{1}{4}Z_F(\chi)F^z_{\mu\nu}F^{z\mu\nu}+iZ_\psi(\chi)
\bar\psi\gamma^\mu D_\mu\psi+\Delta{\cal{L}}\end{equation}
plus additional pieces $\Delta{\cal{L}}$ for the
scalar fields responsible for spontaneous symmetry breaking. Among
them, the Higgs doublet induces masses for the fermions via the Yukawa
couplings. In our normalization\footnote{If the unification
scale is below the Planck mass,
the present value of $f$ is larger than one. This is a pure matter of
convention since we could always
rescale $\chi$ by a constant factor.} $\chi$ equals
the unification scale where
the GUT symmetry is spontaneously broken, $\chi= M_{\rm GUT}$.
The quantity $f$ expresses the ratio between the reduced Planck
mass $\bar M_p=M_p/\sqrt{8\pi}$ and the unification scale
\begin{equation}\label{2.4A}
\frac{\bar M_p}{M_{\rm GUT}}=\frac{f}{\sqrt6}\end{equation}
This ratio may depend on the cosmon field $\chi$ such that in our picture a
time variation of $M_{GUT}/\bar{M}_p$ is described by a nontrivial $\chi$-dependence
of $f(\chi)$. This is equivalent to the more conventional language where $\bar{M}_p$ is
kept fixed and $M_{GUT}$ depends explicitely on time \cite{CFR,LSS}.

We note that ${\cal{L}}_m$ should be associated with the
effective action relevant for momenta $q^2\approx\chi^2$ where the gauge
couplings are unified. The effective action for lower momenta and the
observed particle masses and couplings will be modified by the running of
couplings.
We use a fixed ``bare'' gauge coupling $\bar g$ in the covariant
derivative $D_\mu=\partial_\mu-i\bar g A^z_\mu T_z$ (with $T_z$ appropriate
generators) and in the usual definition of the nonabelian field strength
$F_{\mu\nu}^z$. Then the renormalized gauge coupling $g$ at the scale $\chi$
\begin{equation}\label{D}
g^2(\chi)=\bar g^2Z^{-1}_F(\chi)\end{equation}
depends on $\chi$ through $Z_F(\chi)$. The running of the strong,
weak and electromagnetic gauge couplings for momenta below
$\chi$ is given by the $\beta$-functions  of the standard model. For example,
their values $(\alpha_s=g_s^2/4\pi$ etc.)
at the $\chi$-dependent electroweak scale\footnote{With
$M_W$ the mass of the $W$-boson the gauge hierarchy implies
$\zeta_w\ll 1$. We
will see below that $\zeta_W$ can depend on $\chi$ only weakly \cite{SF}.}
$M_W(\chi)=\zeta_w(\chi)\chi$ are given in the one-loop approximation
and neglecting the (small) contribution of the scalar sector  by
\begin{eqnarray}\label{E}
&&\alpha_s^{-1}(M_W)=\frac{4\pi Z_F(\chi)}{\bar g^2}+\frac{7}{2\pi}\ln
\zeta_w(\chi)\nonumber\\
&&\alpha_w^{-1}(M_W)=\frac{4\pi Z_F(\chi)}{\bar g^2}+\frac{5}{3\pi}\ln
\zeta_w(\chi)\nonumber\\
&&\alpha_{em}^{-1}(M_W)=\frac{32\pi Z_F(\chi)}{3\bar g^2}-\frac{5}{3\pi}\ln
\zeta_w(\chi)\end{eqnarray}
We conclude that the gauge couplings at the weak scale $M_W$ depend on three functions
$Z(\chi),f(\chi)$ and $\zeta_w(\chi)$.
In turn, these functions describe the $\chi$-dependence
(and therefore time dependence) of the mass ratios $\Lambda_{QCD}/M_{GUT},\bar{M}/M_{GUT}$
and $M_W/M_{GUT}$.

\section{Field-dependent mass ratios and fine structure constant}

The nucleon mass (here the neutron mass $m_n$) is in leading
order\footnote{We use here only the three light quarks for the running
of $\alpha_s$ below $M_W$. The corresponding approximation $m_t=m_b=m_c
=M_W$ can easily be improved for a more accurate quantitative
treatment.} proportional to $\Lambda_{\rm QCD}$,
\begin{equation}\label{F}
m_n(\chi)=c_n\Lambda_{\rm QCD}(\chi)=c_nM_W(\chi)\exp\left(
-\frac{2\pi}{9\alpha_s(M_W)}\right)\end{equation}
This implies that the field dependence of the ratio between neutron
mass and Planck mass $(\bar M_p^2=M_p^2/(8\pi)=f^2\chi^2/6)$
obeys
\begin{equation}\label{G}
L_{ng}(\chi)=\ln\frac{m_n(\chi)}{\bar M_p(\chi)}=
\frac{2}{9}\ln\zeta_w-\ln f-\frac{8\pi^2Z_F}{9\bar g^2}+\ {\rm const.}\end{equation}
Similarly, we define the ration
\begin{equation}\label{3.2A}
L_{Wn}(\chi)=\ln\frac{M_W(\chi)}{m_n(\chi)}
\end{equation}
and observe
\begin{equation}\label{GA}
\frac{\partial L_{Wn}}{\partial\ln \chi}=\frac{7}{9}\frac{\partial\ln
\zeta_w}{\partial\ln \chi}+\frac{8\pi^2}{9\bar g^2}\frac{\partial
Z_F}{\partial\ln \chi}\end{equation}
The $\chi$-dependence of $L_{ng}$ can therefore be expressed as
\begin{equation}\label{GB}
\frac{\partial L_{ng}}{\partial\ln\chi}=-\frac{\partial\ln f}{\partial
\ln \chi}-\frac{2\pi}{7\alpha_u}\frac{\partial\ln Z_F}{\partial\ln\chi}+
\frac{2}{7}\frac{\partial L_{Wn}}{\partial\ln\chi}\end{equation}
where we use the GUT-gauge coupling
$\alpha_u=g^2(\chi)/(4\pi)=\bar g^2/(4\pi Z_F(\chi))
\approx 1/40$.
Introducing the shorthands
\begin{eqnarray}\label{GC}
\frac{\partial L_{ng}}{\partial\ln\chi}&=&\beta_{ng}\ ,\quad
\frac{\partial L_{Wn}}{\partial\ln\chi}=\beta_{Wn}\ ,\quad \nonumber\\
\frac{\partial \ln Z_F}{\partial\ln\chi}&=&\eta_F\ ,\quad
\frac{\partial \ln f}{\partial\ln\chi}=B
\end{eqnarray}
we obtain the relation
\begin{equation}\label{GD}
\beta_{ng}=-B-36\eta_F+0.286\beta_{Wn}
\end{equation}

The fine structure constant as observed by transitions in atoms
and molecules corresponds to $\alpha_{em}(m_e)$ and is approximately
given\footnote{We consider again only the effects of the
three light quarks below $M_W$ and approximate $m_\tau\sim m_\mu\sim m_e$,
neglecting $\chi$-independent constants.} by
\begin{eqnarray}\label{H}
\frac{1}{\alpha_{em}(m_e)}&=&\frac{1}{\alpha_{em}(M_W)}+\frac{10}{3\pi}
\ln\frac{M_W}{m_n}+\frac{2}{\pi}\ln
\frac{m_n}{m_e}\\
&=&\frac{368\pi Z_F}{27\bar g^2}+\frac{25}{27\pi}\ln
\zeta_W-\frac{2}{\pi}L_{en}\ +\ {\rm const.} \nonumber
\end{eqnarray}
Here the electron to neutron mass ratio
\begin{equation}\label{I}
L_{en}=\ln\frac{m_e(\chi)}{m_n(\chi)}\end{equation}
typically depends on the Fermi scale and the electron-Yukawa coupling.
The field dependence of the fine structure constant therefore
combines from the $\chi$-dependence of the GUT-gauge coupling $(Z_F)$,
the $\chi$-dependence of the gauge hierarchy $(\zeta_W)$ and $L_{en}$.
Assuming for simplicity $m_e(\chi)\sim M_W(\chi)$ or $L_{en}=L_{Wn}+const$,
the $\chi$-dependence of the fine structure constant becomes
\begin{eqnarray}\label{IA}
\frac{\partial}{\partial\ln\chi}\alpha^{-1}_{em}&=&\frac{22}{7\alpha_u}\frac
{\partial\ln Z_F}{\partial\ln\chi}-\frac{17}{21\pi}\frac{\partial L_{Wn}}
{\partial\ln\chi}\nonumber\\
&=&126\frac{\partial\ln Z_F}{\partial\ln\chi}-
0.26\frac{\partial L_{Wn}}{\partial\ln\chi}
\end{eqnarray}
or
\begin{equation}\label{IB}
\frac{\partial\alpha_{em}}{\partial\ln\chi}=-6.7\cdot10^{-3}\eta_F+1.4
\cdot10^{-5}\beta_{Wn}\end{equation}
Combining eqs. (\ref{GB}) and (\ref{IA}) we can relate the relative
change in the nucleon mass to the relative change in $\alpha_{em}$
\begin{equation}\label{3.10A}
d\ln m_n=\frac{\pi}{11\alpha_{em}}d\ln \alpha_{em}+\frac{7}{33}
d\ln\frac{m_W}{m_n}+d\ln M_{\rm GUT}\end{equation}
This generalizes earlier estimates to situations where the weak scale becomes time
dependent.

Electromagnetic effects and the difference between the
mass of the up and down quark can also induce a $\chi$-dependence in
the ratio between the proton and neutron mass. This will lead to a
composition dependence of gravity-like long-range forces which can be
looked for by tests of the equivalence principle. With \cite{GL}
\begin{eqnarray}\label{J}
m_n-m_p&=&\hat A(m_d-m_u)-\hat B\alpha_{em}m_n,\quad \nonumber\\
\hat A&\approx& 0.6\ ,\quad \hat B\approx0.1
\end{eqnarray}
one finds
\begin{equation}\label{K}
L_{Hn}=\ln\frac{m_p+m_e}{m_n}\approx\hat B\alpha_{em}-\hat A\frac{m_d-m_u}{m_n}+
\frac{m_e}{m_n}
\end{equation}
For constant $\hat A$ and $\hat B$ the $\chi$-dependence of $L_{Hn}$ is
related to the $\chi$-dependence of $\alpha_{em}$ (\ref{IB}) and the
combination
\begin{equation}\label{L}
r_q=\frac{\hat A(m_d-m_u)-m_e}{m_n}=\frac{(\tilde h_1/g_w)M_W}{m_n}\end{equation}
with $\tilde h_1$ a suitable combination of Yukawa couplings.
Neglecting for simplicity the $\chi$-dependence of $\tilde h_1/g_w$ we
may approximate, with $r_q\approx1.3\cdot 10^{-3}$,
\begin{equation}\label{M}
\frac{\partial L_{Hn}}{\partial\ln\chi}\approx 0.1\frac{\partial\alpha_{em}}
{\partial\ln\chi}-r_q\beta_{Wn}\end{equation}
We observe that a contribution $\sim\beta_{Wn}$ will also
appear\footnote{More precisely, $L_{Wn}$ stands here for $\ln(m_q/m_n)$
+\ const., with $m_q$ an appropriate current quark mass.}
in a more refined treatment for which the precise value of $r_q$ in
eq. (\ref{M}) may get somewhat modified.

A second composition dependence can arise from the nuclear binding
energy \cite{PSW}: the relative contribution of the quark mass to the mass
of nuclei may depend on the baryon number B (or also the charge Z). The
average mass of the light quarks,
$m_1=(m_u+m_d)/2$, contributes a fraction
of the nucleon mass\footnote{For these considerations we neglect isospin
violation whose dominant effect is already contained in $L_{Hn}$.}
\begin{equation}\label{P}
m_n=c_n\Lambda_{\rm QCD}+\hat E m_1\end{equation}
Let us introduce an average nuclear binding energy
per nucleon $\epsilon$ and assume, for simplicity,
that it is independent of $m_1$, i.e. $\epsilon \sim \Lambda_{\rm QCD}$.
If $\Lambda_{\rm QCD}$ and $m_1$ depend differently on
$\chi$, the coupling of the cosmon to a nucleus will not be exactly
proportional to mass but  also depend on the relative amount of binding
energy.
One finds, for $m_1\sim M_W$ and  $\sigma\approx 40-60$ MeV the nucleon $\sigma$-term
\cite{GL}:
\begin{eqnarray}\label{Q}
&&\frac{\partial L_{bn}}{\partial\ln\chi}\equiv\frac{\partial
\ln(\epsilon/m_n)}{\partial\ln\chi}\approx -\hat E\frac{\partial(m_1/m_n)}
{\partial\ln\chi}\\
&&\approx-\hat E\frac{m_1}{m_n}\frac{\partial\ln(M_W/m_n)}
{\partial\ln\chi}
=-\frac{\sigma}{m_n}\frac{\partial L_{Wn}}{\partial\ln\chi}\approx-0.05\beta_{Wn}\nonumber
\end{eqnarray}
At this point we have expressed the $\chi$-dependence of all couplings
and mass ratios relevant for our later discussion in terms of three
parameters $\eta_F,\beta_{Wn}$ and $B$. Implicitely we will assume that these parameters
are (approximately) independent of $\chi$ for the cosmological epoch relevant for our
discussion. (This must not hold for the most general scenarios of quintessence.)

\section{Cosmological time variation of the cosmon field}

So far we have investigated the dependence of the fundamental couplings on the cosmon field
$\chi$. In order to translate this into a time dependence we need to know how $\chi$
depends on the the cosmological time $t$. This is the subject of this section. More
explicitely, we need to derive the rate of change of the cosmon field, $d\ln\chi /dt$
for a given model of quintessence. Then we can translate eq. (\ref{IB}) into an equation for
the time dependence of the fine structure constant and similar for other quantities.

In the last two sections we have used a description where all mass
scales depend on $\chi$. This language is useful to underline that
physical observables can only depend on dimensionless couplings or
mass ratios. We never measure a mass or a time difference by itself --
we rather express it in terms of some other mass scale which is used to
set the units. In a strict sense, observables are therefore always
dimensionless quantities. These statements become particularly
apparent once one realizes that one can  change the mass scales
by an appropriate nonlinear field-dependent rescaling of the metric.
For example, we will use below a Weyl scaling such that the Planck
mass becomes a fixed ($\chi$-independent and therefore time-invariant)
unit which we may arbitrarily fix as $10^{19}$ GeV. After the Weyl
scaling, the $\chi$-dependence of a dimensionless ratio as $m_n/\bar M_p$
will be the same as before. However, in the previous language both
$m_n$ and $\bar M_p$ did depend on $\chi$ whereas after the Weyl scaling
only $m_n$ varies with $\chi$ and time. Similarly, there exists a scaling
where $m_n$ becomes a constant and $\bar M_p$ is variable. The choice of the frame of
the metric is a matter of convenience. In a sense, it fixes the
(inverse) mass unit in which time differences are expressed. The cosmological time
evolution  of fundamental couplings is most easily studied in a frame where the Planck
mass is kept fixed.

The Weyl-scaled metric $g_{\mu\nu}$ obtains by multiplying $\bar g_{\mu\nu}$
by a function $w^{-2}(\chi)$ such that
$\bar g_{\mu\nu}=w^2g_{\mu\nu},
\bar g^{1/2}=w^4 g^{1/2}, \bar R=w^{-2}\{R-6(\ln w);^\mu_{\ \mu}-
6(\ln w);^\mu(\ln
w)_{;\mu}\}$. Inserting the particular scaling
function  $w=\sqrt 6\bar M_p/(f\chi)$ the Lagrange density (\ref{B})
reads in the new variables \cite{SF}
\begin{eqnarray}\label{W1}
\sqrt{\bar g}{\cal{L}}_q&=&\sqrt g\{-\frac{1}{2}\bar M^2_pR+\frac{1}{2}
k^2(\varphi)\partial_\mu\varphi\partial^\mu\varphi\nonumber\\
&&+\bar M^4_p
\exp(-\varphi/\bar M_p)\}
\end{eqnarray}
Here $\varphi$ is related to $\chi$ by
\begin{equation}\label{W2}
\varphi=\bar M_p\ln\left(\frac{f^4(\chi)\chi^4}{36 V(\chi)}\right)
\end{equation}
and $k(\varphi)$ is given implicitly as
\begin{equation}\label{W3}
k^2(\varphi)=\frac{1}{16}\delta(\chi)\left(
1+\frac{\partial\ln f}{\partial\ln\chi}-\frac{1}{4}
\frac{\partial\ln V}{\partial\ln\chi}\right)^{-2}\end{equation}
with
\begin{equation}\label{W3A}
\delta(\chi)/6=Z_\chi/f^2+(1+\partial\ln f/\partial\ln\chi)^2\end{equation}
The quantity  $\delta$ measures the deviation from a conformally
invariant kinetic term. We make the important assumption that $f^4
\chi^4/V$ diverges\footnote{Alternatively, $f^4\chi^4/V$ could also
diverge for $\chi\to 0$.} for $\chi\to\infty$ such that
\begin{equation}\label{4.4A}
\lim_{\chi\to\infty}\varphi(\chi)\to\infty\end{equation}
Then one can use a standard exponential form of the
cosmon potential such that the details of the particular model of
quintessence appear in the function $k^2(\varphi)$ which multiplies
the cosmon kinetic term \cite{HW}. We emphasize that the effective action (\ref{W1})
covers all scenarios of quintessence where the kinetic term contains only two derivatives
and the potential decreases monotonically to zero for large values of the scalar field.
Other potentials can be brought to an exponential form by field-rescaling.

We observe that the variable $\varphi$
is well defined as long as $f^4\chi^4/V$ is monotonic in $\chi$ such
that $k(\varphi)$ remains finite. Stability requires $k^2(\varphi)
\geq 0$, which is possible even for negative $Z_\chi$.
For any function $k(\varphi)$ which remains
positive and finite for all $\varphi$
the dynamics of the cosmon-gravity system drives $\varphi$
to infinity at large time. Therefore the cosmological constant
asymptotically vanishes! We emphasize that the condition
(\ref{4.4A}) and the finiteness and
positivity of the expression (\ref{W3}) are sufficient for this
purpose and no fine-tuning of parameters is needed! In particular, adding an
arbitrary constant to $V(\chi)$ does not change this conclusion.

We complete the rescaling by writing the matter and radiation part of the Lagrangian
(\ref{C}) in the frame with constant Planck mass.
Performing also an appropriate Weyl scaling of the spinor fields $\psi\to w^{-3/2}\psi$
and of the Higgs-scalar $H\to w^{-1}H$ (the gauge fields
are not modified), the kinetic terms become
\begin{equation}\label{W6}
\sqrt{\bar g}{\cal{L}}_m=\frac{1}{4}Z_F(\varphi)F^z_{\mu\nu} F^{z\mu\nu}+i
Z_\psi(\varphi)\bar\psi\gamma^\mu D_\mu \psi+...\end{equation}
All mass ratios are invariant with respect to the Weyl scaling.
Therefore the $\chi$-dependence
of the mass ratios directly carries over to a $\varphi$-dependence.
In particular, the quantities $L_{ng},L_{en}$ etc. and the fine
structure constant are not affected by the Weyl scaling.

The time variation of couplings is most conveniently expressed by their
dependence on the redshift $z$ (which is closely related to the cosmological
``look back time''). An estimate needs information about the cosmological history of the
cosmon field. Depending on the question, two pieces of information may be available
for a given model.\\
(i) It may be possible to estimate the difference in the value of $\varphi$ between
some early epoch, say nucleosynthesis, and now, $\delta\varphi(z)=\varphi(z)-\varphi(0)$.
The change of the fine structure constant at the time of nucleosynthesis as compared to
now can then be expressed as
\begin{equation}\label{EX1}
\delta\alpha_{em}(z)=\frac{\partial\alpha_{em}}{\partial \ln\chi}
\frac{\partial\ln\chi}{\partial\varphi}\delta\varphi(z)
\end{equation}
(with straightforward generalization for non-constant $\partial\alpha/\partial\ln\chi$ or
$\partial\ln\chi/\partial\varphi)$.\\
(ii) Differential changes - for example for the recent epoch - can be computed by
\begin{equation}\label{EX2}
\frac{\partial\alpha_{em}(z)}{\partial z}=\frac{\partial\alpha_{em}}{\partial\ln\chi}
\frac{\partial\ln\chi}{\partial z}=\frac{\partial\alpha_{em}}{\partial\ln\chi}G(z)
\end{equation}
We will relate $G(z)$ to the equation of state of quintessence.

In order to compute the r.h.s. of eqs. (\ref{EX1})(\ref{EX2}) we have to trace back
the cosmological history of $\varphi$. A useful form of the cosmological evolution
equation \cite{HW} for $\varphi$ replaces time by $\ln a=-\ln(1+z)$, i.e.
\begin{eqnarray}\label{AB0}
\frac{d\varphi}{d\ln a}&=&\bar{M}_p\left(\frac{6(\rho_{\varphi}-V(\varphi))}
{k^2(\varphi)(\rho_m+\rho_r+\rho_{\varphi}}\right)^{1/2}\\
\frac{d\ln\rho_{\varphi}}{d\ln a}&=&-6\left(1-\frac{V(\varphi)}{\rho_{\varphi}}
\right)\label{AB1}
\end{eqnarray}
Here $V(\varphi)=\bar{M}_p^4\exp(-\varphi/\bar{M}_p)~,~\rho_{\varphi}=V(\varphi)+
\dot{\varphi}^2 /2$ is the total energy density of quintessence, and $\rho_m,\rho_{\rho}$
stand for the energy density of matter and radiation. We assume zero spatial curvative
such that $\rho_m+\rho_r+\rho_{\varphi}=\rho_{cr}=3\bar{M}^2_pH^2=3\bar{M}^2_p
(\dot{a}/a)^2$. The solution to these equations yields directly the dependence
of $\varphi$ on the redshift $z$, and, via eq. (\ref{W2}), therefore $d\ln\chi/dz$.

To be more specific, let us consider a behavior for large $\chi$ with
$V\sim \chi^{4-A},\ f\sim \chi^B$ such that
\begin{equation}\label{W4}
k^2=\frac{\delta}{(A+4B)^2}
\end{equation}
This can easily be extended to the most general cosmon potential and gravitational
coupling if we allow A and B to depend on $\chi$
\begin{equation}\label{4.6AA}
A=4-\frac{\partial\ln V}{\partial\ln\chi}
\end{equation}
For small enough and slowly varying $k(\varphi)$ the cosmology
of our model is well known \cite{CW2}. It behaves
like ``exponential quintessence'' where the
fraction of dark energy approaches a constant value.
During the radiation-dominated $(n_b=4)$ or
matter-dominated $(n_b=3)$ universe, the homogeneous quintessence $(\Omega_h=\rho_{\varphi}
/\rho_{cr})$ reaches after some ``initial evolution'' the value
\begin{equation}\label{W5}
\Omega_h=n_bk^2
\end{equation}
This is independent of the precise initial condition.
A small amount of dark energy in early cosmology typically requires
approximate conformal symmetry of the kinetic term, $\delta \ll 1$. In a more
recent epoch, however, the qualitative behavior may have changed. An increase
of $\delta$ (and therefore $k$) can lead to a realistic cosmology where
$\rho_{\varphi}$ accounts today for two third of the total energy density,
$\Omega_h\approx 2/3$ \cite{HW,CWCF}.

Our conventions have the advantage that the value of $\varphi$ at some earlier epoch in
cosmology is directly related to the potential energy of the
scalar field at a given redshift
\begin{equation}\label{T1}
V(z)=\bar M^4_p\exp\left(-\frac{\varphi(z)}
{\bar M_p}\right)\end{equation}
The difference  between the value $\varphi(z)$ in some
earlier epoch and today's cosmological
value $\varphi_0$ obeys therefore the simple relation
\begin{equation}\label{T2}
\delta\varphi(z)=\varphi(z)-\varphi(0)=-\bar M_p\ln\left(\frac
{V(z)}{V(0)}\right)\end{equation}
The corresponding value of $\chi$ follows from eq. (\ref{W2})
and we observe
\begin{eqnarray}\label{T2A}
\frac{\partial\ln\chi}{\partial\varphi}
&=&[4\bar M_p(1+\frac{\partial\ln f}{\partial\ln \chi}-\frac{1}{4}
\frac{\partial\ln V}{\partial\ln\chi})]^{-1}\nonumber\\
&=&\frac{k}{\bar M_p\sqrt{\delta}}=\frac{1}{(A+4B)\bar M_p}
\end{eqnarray}
In our conventions $k$ is positive if $\varphi$ is a
monotonically increasing function of $\chi$. With the help
of the relations (\ref{T2}) and (\ref{T2A}) we can now directly
transfer the $\chi$-dependence of couplings and mass ratios
computed in the last section to a dependence on $\varphi$ and
therefore on redshift.
\begin{equation}\label{4.11ZZ}
\delta\alpha_{em}(z)=\frac{1}{A+4B}\frac{\partial\alpha_{em}}{\partial\ln\chi}
\frac{\delta\varphi(z)}{M_p}
\end{equation}
At this point $A$ enters as an additional
parameter. For general quintessence models $k(\varphi)$ determines
the cosmology and therefore $V(z)$ whereas $\delta(\varphi)$ is
needed for the relation between cosmology and time-varying couplings.

We next want to estimate the differential redshift dependence (\ref{EX2}) by
computing $G(z)=\partial\ln\chi/\partial z$. One useful form combines eqs. (\ref{T2A})
and (\ref{AB0}),
\begin{equation}\label{AB3}
G(z)=\frac{\partial\ln\chi}{\partial z}=-\frac{1}{1+z}
\left(\frac{3\Omega_h(1+w_h)}{\delta}\right)^{1/2}
\end{equation}
Here we use the equation of the state of quintessence
\begin{equation}\label{AB4}
w_h=\frac{p_{\varphi}}{\rho_{\varphi}}=\frac{T-V}{T+V}~,~
T=\frac{1}{2}\dot{\varphi}^2
\end{equation}
We observe that the time variation of couplings is suppressed for large $\delta$
and for $w_h\rightarrow -1$. In particular, there exist models of quintessence where the
time evolution of quintessence is very slow today such that $T\ll|V|~,~w_h\rightarrow -1$.
Then the phenomenology approaches the one for a cosmological constant for which no time
variation of the couplings is expected.

Another useful relation expresses $G(z)$ in terms of $w_h(z)$. For this purpose we invert
eq. (\ref{AB0})
\begin{eqnarray}\label{AB5}
k(\varphi)&=&-\big(3\Omega_h(1+w_h)\big)^{1/2}/(d\ln V/d\ln a)\nonumber\\
&=&\frac{\sqrt{3\Omega_h(1+w_h)}}{3(1+w_h)-d\ln(1-w_h)/d\ln a}
\end{eqnarray}
and use $V/\rho_{\varphi}=(1-w_h)/2$. Inserting this in eqs. (\ref{AB3}), (\ref{T2A})
yields the wanted result
\begin{equation}\label{AB7}
G(z)=\frac{\partial\ln\chi}{\partial z}=-\frac{1}{A+4B}
\left\{\frac {3(1+w_h)}{1+z}-\frac{dw_h/dz}{1-w_h}\right\}
\end{equation}

We can draw a few simple conclusions:\\
(i) The quantities $\eta_F~,~\beta_{Wn}$ and $B$ introduced in the last two sections
always appear in the combinations $\eta_F/(A+4B)~,~\beta_{Wn}/(A+4B)~,~B/(A+4B)$ as
far as the time evolution is concerned.\\
(ii) The dependence of the fundamental constants on $z$ is not linear.
Even for $\eta_F,\beta_{Wn},A,B$ and $w_h$ constant the rate of change of the couplings
decreases for large $z$ proportional $(1+z)^{-1}$.\\
(iii) The change in $\alpha_{em}$ is reduced during cosmological epochs when $w_h$
decreases with decreasing $z$ or when $w_h$ comes close to $-1$. This is what happens in
``leaping kinetic term quintessence'' \cite{HW} or ``conformal quintessence''
\cite{CWCF} in the very recent period. Such a feature may explain why observations seem to
imply that $d\alpha_{em}/dz$ is much smaller for $z\approx 0$ than for $z\approx 2$.\\
(iv) We finally observe that $\Omega_h$ does not appear in the relation between $\chi$
and $z$ and therefore does not affect the time evolution of the fundamental couplings.
The latter depends only on the equation of state $w_h$. Of course, time varying
couplings are possible even if the cosmon plays no role in late cosmology, i.e.
$\Omega_h\ll 1$.

\section{Time dependence of the fine-structure constant}

In this section we turn to the time variation of the fine constant as reported
\cite{Obs} from quasar absorption lines at redshift $z\approx 2$.
For $z\approx2$ the recent history of quintessence matters
and $\delta\varphi$
depends sensitively on the particular model. We take here as a
reasonable value for
this range $\delta\varphi(z)/\bar M_p\approx-2$. This means that
the potential energy of the cosmon field at $z\approx 2$ was about
eight times its present value.
We can now estimate the relative change in $\alpha_{em}$ at
$z\approx2$ as compared to today
\begin{eqnarray}\label{T3}
\frac{\delta\alpha_{em}}{\alpha_{em}}&\approx&\frac{\delta\varphi}{\alpha_{em}}
\frac{\partial\alpha_{em}}{\partial\varphi}\approx 2\bar M_p\alpha_{em}
\frac{\partial\alpha_{em}^{-1}}{\partial\ln \chi}\frac{\partial\ln\chi}
{\partial\varphi}\nonumber\\
&=&\alpha_{em}\left(\frac{11}{7\alpha_u}\frac{\partial\ln Z_F}{\partial
\ln \chi}-\frac{17}{42\pi}\frac{\partial L_{Wn}}{\partial\ln\chi}\right)\nonumber\\
&&\hspace{1cm}\left(1+\frac{\partial\ln f}{\partial \ln\chi}-\frac{1}{4}
\frac{\partial\ln V}{\partial\ln \chi}\right)^{-1}
\end{eqnarray}
Using the definitions (\ref{GC}) (\ref{4.6AA}) we arrive at the result
that at redshifts $z\approx2$ the relative change in the fine structure
constant is given by
\begin{equation}\label{T4}
\frac{\delta\alpha_{em}}{\alpha_{em}}\approx(1.84 \eta_F+3.8\cdot 10^{-3}
\beta_{Wn})/(A+4B)\end{equation}
This is to be compared with the reported measurement \cite{Obs}
$\delta\alpha
_{em}/\alpha_{em}\approx-0.7\cdot10^{-5}$. If the effect is mainly
due to the $\chi$-dependence  of $Z_F$ we conclude
\begin{equation}\label{T5}
\eta_F\approx-3.8\cdot 10^{-6}(A+4B)\end{equation}
On the other hand, if the $\chi$-dependence of the ratio between
the weak and strong scales dominates, one finds
\begin{equation}\label{T6}
\beta_{Wn}\approx-1.8\cdot10^{-3}(A+4B)\end{equation}
We assume here that there is no accidental cancellation between the
two contributions such that either (\ref{T5}) or (\ref{T6})
(or both) should hold approximately.

\section{Modifications of nucleosynthesis}

Let us next turn to nucleosynthesis $(z\approx 10^{10})$. Assuming
the presence of early quintessence at the few percent level,
we can estimate $V_{ns}\approx 10^{-2}\rho_{ns}\approx3\cdot 10^{-2}
\bar M_p^2H^2_{ns}$ or, with $V_0\approx\rho_0$,
\begin{eqnarray}\label{T7}
\delta\varphi_{ns}&=&-2\bar M_p(\ln(H_{ns}/H_0)-\ln10)\nonumber\\
&\approx&-2\bar M_p\ln\left(\frac{t_0}{10t_{ns}}\right)\approx-75\bar M_p\end{eqnarray}
We consider essentially constant
$\partial\ln V/\partial\ln\chi=4-A,\ \partial
\ln f/\partial\ln\chi=B$ such that the corresponding value of $\chi$ obeys
\begin{equation}\label{T8}
\ln\left(\frac{\chi_{ns}}{\chi_0}\right)=-\frac{75}{A+4B}\end{equation}
If we further assume that $\eta_F$ and $\beta_{Wn}$ are almost $\chi$-independent,
this yields
(cf. eq. (\ref{IA})) for the fine-structure constant during
nucleosynthesis
\begin{equation}\label{T9}
\frac{1}{\alpha_{em}^{ns}}-\frac{1}{\alpha^0_{em}}=-\frac{0.95\cdot
10^4}{A+4B}\eta_F-\frac{19.5}{A+4B}\beta_{Wn}
\approx0.035\end{equation}
The relative change $\Delta\ln\alpha_{em}\approx -2.6\cdot 10^{-4}$ is
presumably too small to be observable at present by purely electromagnetic effects
\cite{NSA}, \cite{AV}. The same holds for the cosmic microwave background \cite{Mar}
since the relative change in $\alpha_{em}$ at last scattering is smaller
\footnote{We discuss here the homogeneous spatially averaged value of $\alpha_{em}$
and define the relative change in comparison with the present value.
For a discussion of inhomogeneities in $\alpha_{em}$ see \cite{Bar}.}
than for nucleosynthesis if $\alpha_{em}$ depends monotonically on $\chi$.

However, in a grand unified theory the change in $\alpha_{em}$ is related to the change
of the nucleon mass by eq. (\ref{3.10A}). The variation of $m_n/\bar{M}_p$ constitutes the
dominant effect of time varying couplings for nucleosynthesis \cite{SF,CW3,NSA}. The
analogue of eq. (\ref{T9}) for the nucleon mass (for fixed Planck mass)
obtains from eq. (\ref{GB})
\begin{equation}\label{T10}
\frac{m^{ns}_n}{m^0_n}=\exp\left\{\frac{75B}{A+4B}+\frac{2700}{A+4B}
\eta_F-\frac{21.4}{A+4B}\beta_{Wn}\right\}\end{equation}
The relative change due to the $\chi$-dependence of $Z_F$ is
$\approx 10^{-2}$. There has been a debate \cite{NSA} about the precision with which such a
change can be seen with the present accuracy of element abundances. (Remember
that this quantity concerns the common mass scale for neutrons and protons.)
If one takes the attitude that the observed He-abundance deviates by no more than $0.8 \%$
from the value computed with constant couplings one infers the bound
$\Delta\ln(m_n/\bar{M}_p)<0.025$. (This corresponds to $\Delta\ln\alpha_{em}<6.4\cdot 10^{-4}$
if $\eta_F$ dominates.) We find that this bound is obeyed if $\Delta\alpha_{em}$ is
dominated by $\eta_F$.

On the other hand, if the time-dependence of $\alpha_{em}$ is due
to the change in $L_{Wn}$, the relative change in the  nucleon mass
$\delta m_n/m_n\approx 4\cdot 10^{-2}$ turns out to be substantially larger.
This may indeed influence the details
of nucleosynthesis. Inserting the value (\ref{T6}), all QCD mass scales
(including binding energies etc.) would be enhanced by 4 \% as compared
to today. Equivalently, we may keep a fixed strong interaction
scale\footnote{Furthermore, for $\beta_{Wn}\not=0$ there are effects
from the change of weak interaction and decay rates as compared
to a fixed nucleon mass which are discussed below.}
and discuss the effect of a Planck mass that is lower by 4 \% as compared
to today. (Only the ratio matters!) This would affect the relation
between the temperature and the Hubble parameter and therefore modify the
relevant ``clock''. The net effect is a value of $H^2$ that is larger
by 8 \% for a given temperature characteristic for nucleosynthesis, similar
to the addition of half a neutrino species.

An important constraint arises for the $\chi$-dependence of $f$.
Discarding conservatively a change of more than 10 \% of the nucleon mass at
nucleosynthesis, we obtain the bound
\begin{equation}\label{T11}
|\frac{B}{A+4B}|\stackrel{\scriptstyle<}{\sim} 1.3\cdot 10^{-3}
\end{equation}
For $\Delta\ln(m_n/\bar{M}_p)<0.025$ as discussed above this bound gets more
severe by a factor 4.

We note, however, that for negative $B$ the
ratio $m_n/\bar M_p$ during nucleosynthesis would be lower than
in the usual scenario. Therefore the effect of the reduced
nucleon mass (or enhanced Planck mass)
could be compensated by a larger amount of quintessence during
nucleosynthesis. In order to test possible effects of this type, we
propose to treat the proportionality constant in the law $H\sim T^2$
as a free parameter. The number of ``effective neutrino species''
could come out even smaller than three. Such an outcome would
point to a time variation of fundamental ``constants''!

We also may have a look at the change in characteristic mass ratios or
interaction rates during nucleosynthesis. The first concerns the ratio
$m_e/m_n$ which is given in our approach by $\exp L_{Wn}$. With
eq. (\ref{T6}) this ratio would increase by 14 \%!
Similarly, the pion to nucleon mass ratio would increase by
7 \%. (In our approximation $m_u+m_d\sim m_e$ one has
$\delta\ln(m_\pi/m_n)=\frac{1}{2}\delta\ln(m_e/m_n)$). The changes
in weak interaction rates $\sim M^{-4}_W$ are even more dramatic:
the $\beta$-decay rate of the neutron would decrease by more than
50 \%! This effect seems too strong to be acceptable, and we conclude
that the nucleosynthesis bound on the size of $\beta_{Wn}$
seems to be more severe
than the estimate (\ref{T6}). In turn, this suggests that the most
plausible origin of a time dependence in $\alpha_{em}$ arises from the
$\chi$-dependence of $Z_F$. A more detailed analysis would
be welcome. We will see in the next section that even more restrictive
bounds on $\beta_{Wn}$ arise from tests of the equivalence principle.

Finally, we briefly discuss the proton-to-neutron mass ratio
which obeys a formula closely similar to (\ref{M}). The
electromagnetic effect is tiny (a few times $10^{-5}$) and
we concentrate on the change of $L_{Wn}$. The relative change
for nucleosynthesis is
\begin{equation}\label{T12}
\delta\ln\frac{m_p}{m_n}\approx\frac{0.1}
{A+4B}\beta_{Wn}\end{equation}
This  corresponds to $2\cdot 10^{-4}$ inserting
(\ref{T6}) and becomes even smaller for smaller
$|\beta_{Wn}|$. We conclude that the most important issues for
nucleosynthesis are the possible change of the weak interaction
rates which constrains $\beta_{Wn}$ and the
change in the clock which constrains $B$.

\section{Composition-dependent forces with gravitational strength}

If the time variation of the fine structure constant is explained by the
variation of a scalar field this immediately implies the existence of a new interaction.
The cosmon field can depend on time and on space, and the space variation is intimately
connected to a force. Similar to gravity or electromagnetism a source for the cosmon field
will generate an inhomogeneous cosmon field around it. For a test body in the cosmon field
the potential energy depends on the location, resulting in an effective force. For the
cosmon, this force is long ranged and composition dependent \cite{PSW}.

In this section we discuss the composition dependence of the
``fifth force''-type interaction mediated by the exchange of the
cosmon. This will yield restrictive bounds for the cosmon coupling to matter and radiation.

The cosmon-interaction is most easily computed in the Weyl-scaled language
of sect. 4, where the kinetic terms for the cosmon and graviton are diagonal. On
scales of our solar system or smaller the cosmon is massless
for all practical purposes \footnote{In this respect quintessence differs from the
intermediate range force discussed in \cite{PSW}.}. Its coupling to matter is at most
of gravitational strength. It is therefore convenient to
interpret the effects of the cosmon exchange as modifications
of Newtonian gravity which are dependent on the composition of
the test bodies. For a fermion at rest the gravitational charge is
given by its renormalized mass $m_f/(\sqrt2\bar M_p)$, whereas
the corresponding cosmon charge $Q_f$ obtains
by linearizing the $\varphi$-dependence of the mass
$m_f=m_f(\varphi_0)+(\partial m/\partial \varphi)(\varphi_0)
(\varphi-\varphi_0)$ as
\begin{equation}\label{W7}
Q_f=k^{-1}\frac{\partial m_f}{\partial\varphi}
\end{equation}
Here the factor $k^{-1}$ reflects that a field $\tilde\varphi$ with
standard kinetic term is related to $\varphi$ by $d\tilde\varphi=kd\varphi$.
In complete analogy to gravity the cosmon mediates a composition-dependent correction
to the gravitational $1/r$ potential. Newton's law
for the attraction between to identical fermions is multiplied
by a factor $(1+\alpha_f)$
\begin{eqnarray}\label{W8}
V_N&=&-\frac{G_Nm_f^2}{r}(1+\alpha_f),\nonumber\\
\alpha_f&=&\frac{2Q_f^2\bar M^2_p}{m_f^2}
=\frac{2\bar M^2_p}{k^2}\left(\frac{\partial\ln m_f}{\partial\varphi}
\right)^2
\end{eqnarray}
In eq. (\ref{W8}) the quantities $k$ and $\partial\ln m_f/d\varphi$ and therefore
$\alpha_f$ are evaluated at the appropriate ``background value''
$\varphi_0$. The composition dependence arises from the fact that
$\partial\ln m_f/\partial\varphi$ is different for different species of fermions.

This discussion is easily generalized to a situation with many particles. For
laboratory experiments one computes the total cosmon charge of the earth, $Q_E=k^{-1}
\partial M_E/\partial\varphi$, in complete analogy to the total mass $M_E$ relevant for
gravity. This will then get multiplied by individual cosmon charges of the test bodies
(\ref{W7}). Due to the very long range of the cosmon interaction the acceleration
of the test bodies does not depend on the location - this differs from earlier versions
of a scalar mediated fifth force
with intermediate range \cite{PSW}. The relative acceleration
of the test bodies depends solely on their composition. In turn, the cosmon charge can be
related to the cosmon coupling to matter and radiation discussed in sects. 2, 3 by
\begin{equation}\label{5.AA1}
\frac{\bar{M}_p}{k}\frac{\partial\ln m}{\partial\varphi}=
\frac{1}{k(A+4B)}\frac{\partial\ln m}{\partial\ln\chi}
\end{equation}
We notice that the normalization factor $k^{-1}$ implies that bounds from tests of the
equivalence principle will always involve products of the type
$\eta_F/\big((A+4B)k\big)$, in contrast to bounds from the time variation of couplings
where the factor $k^{-1}$ does not appear explicitely. This seems to make the relation
somewhat less direct. We emphasize, however, that $k(\varphi_0)$ is closely related
to the equation of state of quintessence at the present epoch according to eq.
(\ref{AB5}). At the end of this section we will present an explicit relation between
$\partial\alpha_{em}/\partial z$ at $z=0$ and the tests of the equivalence principle.

We want to compute the relative differential acceleration $\eta=
2|a_1-a_2|/|a_1+a_2|$ for two test bodies with equal mass $m_t$
but different composition. For this purpose we describe the earth as
a collection of $N_E$ neutrons and $Z_E$ protons, and similar for
the test bodies. The total mass of an atom and similar for extended
objects is given by $(B=N+Z)$
\begin{equation}\label{W9}
M=Nm_n+Zm_H+B\epsilon
\end{equation}
with $m_H=m_p+m_e$ and $\epsilon$ the nuclear binding energy per
baryon\footnote{The proportionality of the nuclear binding
energy to baryon number is an approximation that could be replaced
by a more complicated nuclear mass formula.}.
This yields
\begin{equation}\label{W10}
\eta=\frac{2\bar M^2_p}{k^2M_Em_t}\frac{\partial M_E}{\partial\varphi}
\left(\Delta N\frac{\partial m_n}{\partial\varphi}+\Delta Z
\frac{\partial m_H}{\partial\varphi}+\Delta B\frac{\partial\epsilon}
{\partial\varphi}\right)\end{equation}
where the equality of the test masses implies
\begin{equation}\label{W11}
\Delta Nm_n+\Delta Zm_H+\Delta B\epsilon=0\end{equation}
We also define $\bar m=M_E/(N_E+Z_E)$, the difference in the proton
fraction of the two test bodies $\Delta R_Z=\Delta Z m_H/m_t$
and similar for the binding energy $\Delta R_B=\Delta B\epsilon/
m_t$. The differential acceleration
\begin{equation}\label{W12}
\eta=\frac{2\bar M_p^2}{k^2}\frac{\partial\ln\bar m}{\partial
\varphi}\left(\Delta R_Z\frac{\partial L_{Hn}}{\partial\varphi}
+\Delta R_B\frac{\partial L_{bn}}{\partial\varphi}\right)\end{equation}
can then be related
to the $\varphi$-dependence of the partial mass ratios
as given by $L_{Hn}$ and $L_{bn}$ in eqs. (\ref{K}), (\ref{Q}). Typical experimental
bounds \cite{EOT} $|\eta|\leq 3\cdot 10^{-13}$ are obtained for materials
with $\Delta R_Z\approx 0.06-0.1,\ \Delta R_B\approx 3\cdot
10^{-3}$ or smaller.

With eq. (\ref{T2A}) and $\partial\ln\bar m/\partial\varphi\approx
(\partial L_{ng}/\partial\ln\chi)$
$(\partial\ln\chi/\partial\varphi)$
we can finally express $\eta$ as a function of the quantities estimated
previously (\ref{GB}), (\ref{M}), (\ref{Q})
\begin{equation}\label{W14}
\eta=\frac{2}{\delta}\frac{\partial L_{ng}}{\partial\ln\chi}
\left(\Delta R_Z\frac{\partial L_{Hn}}{\partial\ln\chi}+
\Delta R_B\frac{\partial L_{bn}}{\partial\ln\chi}\right)
\end{equation}
and therefore relate it to the functions $f, V, Z_\chi$ an $Z_F$
introduced in eqs. (\ref{B}), (\ref{C}). Since comparison
of different materials is used in the tests for composition-dependent
forces, we exclude cancellations and apply the bound for $\eta$ separately
to the two contributions in the bracket of eq. (\ref{W14}).

We consider first the case where the dominant contribution arises from
the $\chi$-dependence of $Z_F$ and is given by eq. (\ref{T5}). This yields
(in the limit $\beta_{Wn}/\eta_F\to 0,\ B/\eta_F\to 0$)
\begin{eqnarray}\label{W15}
\frac{\partial L_{Hn}}{\partial\ln\chi}&=&0.1\frac{\partial\alpha_{em}}{
\partial\ln\chi}\approx2.5\cdot 10^{-9}(A+4B)\quad ,\quad \frac{\partial L_{bn}}
{\partial\ln\chi}=0,\nonumber\\
\frac{\partial L_{ng}}{\partial\ln\chi}&=&-36\eta_F=1.4\cdot10^{-4}
(A+4B)\end{eqnarray}
and one estimates
\begin{equation}\label{W16}
\eta\approx-7\cdot 10^{-14}/k^2\end{equation}
This comes close to the present experimental bounds! The issue
depends crucially on the value of $k^2$. In early cosmology the
amount of quintessence cannot have been too important. For
$\Omega_d<0.2$ during structure formation one has $k^2<0.07$
during this period (cf. eq. (\ref{W5})). For this range of $k^2$
one would expect that a violation of the equivalence principle
should already have been detected. However, if quintessence accounts
for a large fraction of dark energy today, the
effective value of $k^2$ must be higher at present \cite{HW},
typically of the order one. For these larger values
\footnote{It is not excluded that the microscopic
value of $k^2$ which enters eq. (\ref{W16}) differs from the effective
value of $k^2$ which determines the time evolution of quintessence. It
is conceivable that the present comparably large value of $k^2$
is an effect of the
``backreaction'' of structure formation \cite{CWBR}, whereas
the microscopic value is smaller.} of $k^2$ we still
infer that composition-dependent modifications of Newton's law
should be in the detectable range!

We conclude that the compatibility of the reported time variation
of the fine structure constant with present bounds on composition-dependent
modifications of gravity starts to place bounds on particular models
of quintessence. With $k^2>0.7$ and constant $w_h$ we infer from eq. (\ref{AB5})
the bound
\begin{equation}\label{7.9XX}
w_h\stackrel{<}{\sim}-0.65
\end{equation}
This bound is comparable to the one from supernovae observations \cite{SN}.
It relies, of course, on the assumption that the reported variation of the fine
structure is confirmed. We emphasize that our estimate of the strength of
composition-dependent forces is much higher than in refs. \cite{DZ}, \cite{CK}.
This is due to the fact that not only purely electromagnetic effects
enter into the determination of the cosmon field of the earth. The
latter is proportional to $\beta_{ng}$, and in eq. (\ref{GB})
we  relate this quantity to the
$\chi$-dependence of the ratio between the strong interaction
scale and the Planck mass.

The strength of the source of the cosmon field can be
enhanced by a larger value of $\partial L_{ng}/\partial\ln \chi$ due to
a nonvanishing $B=\partial\ln f/\partial\ln \chi$, cf. eq. (\ref{GD}). If we insert
the upper bound (\ref{T11}), this would enhance $\eta$ by a factor
of 10. Clearly, this brings the differential acceleration into
a range where it should have been observed
by experiment! The present bounds on $\eta$ may therefore
be used to strengthen the bound on $|B|$ as compared
to the bound (\ref{T11}) from nucleosynthesis if $k^2<2$, i.e.
\begin{equation}\label{W17}
\left|\frac{B}{A+4B}\right|\stackrel{\scriptstyle<}{\sim}
6\cdot 10^{-4}k^2
\end{equation}

Finally, we look at the possibility that the composition-dependent
force is dominated by the $\chi$-dependence of
the ratio between the weak and strong
interaction scales $\partial L_{Wn}/\partial\ln\chi$. In this case,
one has
\begin{eqnarray}\label{W17a}
\frac{\partial L_{Hn}}{\partial\ln\chi}&=&-1.3\beta_{Wn},\quad
\frac{\partial L_{bn}}{\partial \ln\chi}=-0.05\beta_{Wn},\nonumber\\
\quad\frac{\partial L_{ng}}{\partial\ln\chi}&=&0.286\beta_{Wn}
\end{eqnarray}
and the contribution from the $\chi$-dependence of the binding
energy and the proton-neutron mass difference are of the same
order. We conclude
\begin{equation}\label{W18}
|\eta|=\frac{8.6\cdot 10^{-5}}{k^2}[\beta_{Wn}/(A+4B)]^2\end{equation}
and infer the approximate bound
\begin{equation}\label{W19}
|\beta_{Wn}|\stackrel{\scriptstyle<}{\sim} 6 \cdot 10^{-5}k(A+4B)\end{equation}
This is so far the strongest bound on $\beta_{Wn}$. Comparing with eq.
(\ref{T6}), we see that the $\chi$-dependence of $L_{Wn}$ can only give
a small contribution to the time dependence of the fine structure
constant and may be neglected!

Furthermore, the bound may even get
stronger if $\beta_{ng}$ is dominated by
$\eta_F$ or $B$. Indeed, we learn from eq. (\ref{W19}) that the contribution
$\sim\beta_{Wn}$ is typically smaller than the first two terms in eq.
(\ref{GD}). If $\beta_{ng}$ is dominated by $\eta_F$, the limit on
composition-dependent forces implies
\begin{equation}\label{7.14}
|\beta_{Wn}|\stackrel{\scriptstyle<}{\sim} 7\cdot 10^{-6}k^2|A+4B|
\end{equation}
This bound holds unless the first two terms in eq. (\ref{GD}) cancel each
other. The small value of $|\beta_{Wn}|$ implies that the second term on
the r.h.s. of eq. (\ref{3.10A}) is small such that
\begin{equation}\label{7.15}
d\ln\frac{m_n}{M_{\rm GUT}}=\frac{\pi}{11\alpha_{em}}d\ln\alpha_{em}=39 d\ln
\alpha_{em}\end{equation}
This relates the variation of the fine structure constant to
the variation of the nucleon mass. Our formula (\ref{7.15}) is somewhat
analogous to \cite{CFR}. We emphasize, however, that it involves
$M_{\rm GUT}$ and not $\bar M_p$. This is important since the appearance
of $B$ in
\begin{eqnarray}\label{7.16}
d\ln\frac{m_n}{\bar M_p}&=&d\ln\frac{m_n}{M_{\rm GUT}} +
d\ln\frac{M_{\rm GUT}}{\bar M_p}\nonumber\\
&=&\frac{\pi}{11\alpha_{em}}d\ln\alpha_{em}-Bd\ln\chi\end{eqnarray}
destroys the one to one correspondence between the variations
of the two quantities.

In the limit $B=0,\beta_{Wn}=0$ the differential acceleration can be related directly to
$\partial\alpha_{em}/\partial zs$ at zero redshift. Indeed, with eqs. (\ref{EX2})(\ref{AB4})
one has
\begin{equation}\label{X11A}
\frac{\partial\ln\alpha_{em}(z)}{\partial z}_{|z=0}=
\eta_F G(0)=-\eta_F
\left(\frac{3\Omega^{(0)}_h(1+w^{(0)}_h)}{\delta}\right)^{1/2}
\end{equation}
Eq. (\ref{W14}) yields
\begin{equation}\label{X11B}
\eta=-7.2\alpha_{em}\Delta R_Z\eta^2_F/\delta
\end{equation}
and we observe that the relation
\begin{equation}\label{X11C}
\eta=-1.75\cdot 10^{-2}
\left(\frac{\partial\ln\alpha_{em}}{\partial z}\right)^2_{|z=0}
\frac{\Delta R_Z}{\Omega^{(0)}_h(1+w^{(0)}_h)}
\end{equation}
becomes independent of $k$ or $\delta$. In this relation the only model dependence appears
through the quantities $\Omega^{(0)}_h$ and $w^{(0)}_h$. We emphasize that the relation
(\ref{X11C}) holds for arbitrary scalar fields mediating the new force. This can be traced
back to the observation that the expression
\begin{equation}\label{X11D}
\frac{\partial\ln\alpha_{em}}{\partial z}=-
\frac{\partial\ln\alpha_{em}}{\partial\tilde{\varphi}}\frac{\dot{\tilde{\varphi}}}{H}
\end{equation}
is independent of the normalization of $\tilde{\varphi}$. We can therefore choose a standard
kinetic term with $T=\dot{\tilde{\varphi}}^2/2=\Omega_h(1+w_h)\rho_{cr}/2$ or
\begin{equation}\label{X11E}
\frac{\dot{\tilde{\varphi}}^2}{H^2}=3\Omega_h(1+w_h).
\end{equation}
Therefore the scalar charge Q (eq. (\ref{W7}))
\begin{eqnarray}
Q&=&\frac{\partial m}{\partial\tilde{\varphi}}=\frac{\partial m}{\partial\ln\alpha_{em}}
\frac{\partial\ln\alpha_{em}}{\partial\tilde{\varphi}}\nonumber\\
&=&-\frac{\partial m}{\partial\ln\alpha_{em}}\frac{\partial\ln\alpha_{em}}{\partial z}
\big[3\Omega_h(1+w_h)\big]^{-1/2}
\end{eqnarray}
involves besides $\partial\ln\alpha_{em}/\partial z$ (at $z=0$) only the dependence
of the masses on $\alpha_{em}$ (including the indirect influence by eq. (\ref{7.16})) and
the fraction of scalar energy density over the critical energy density $\Omega_h$, as well
as the equation of state $w_h$. Since $1+w_h\leq 2$ a sizeable variation of fundamental
constants is compatible with the bounds on the differential acceleration only if
$\Omega_h$ is not too far below $1$! This provides an additional argument in favor of the
association of the scalar field with the cosmon. Then $\Omega^{(0)}_h\approx 0.7$ and
$w^{(0)}_h$ can be measured by cosmological observations.

The factor $\tilde{Q}$ in eq. (\ref{1BB}) accounts for the effects of $\beta_{Wn}$ and $B$
that we have discussed quantitatively above.

\section{Bounds on field-dependent couplings}

We have seen that tests of the time variation of the fine structure
constant, cosmological constraints from nucleosynthesis and the search
for composition-dependent corrections to Newton's law all test the
field dependence of couplings. We have simplified the situation by
concentrating on four quantities, $\eta_F=\partial\ln Z_F/\partial\ln\chi,
\ B=\partial\ln f/\partial\ln\chi$,
$A=4-\partial\ln V/\partial\ln\chi$ and $\beta_{Wn}=
\partial \ln(m_W/m_n)/\partial\ln\chi$. The first three are directly
related to the effective action at the unification scale
(\ref{B}), (\ref{C}), whereas $\beta_{Wn}$ stands collectively
for the $\chi$-dependence of various couplings relevant for weak
interactions and fermion masses.
The first quantity,
\begin{equation}\label{B1}
\eta_F=\frac{\partial\ln Z_F}{\partial\ln\chi}=-\frac{1}{2g(\chi)}\frac{
\partial g(\chi)}{\partial\ln\chi}=-\frac{1}{2}\frac{\beta_g}{g}\end{equation}
describes the dependence of the gauge coupling on the unification scale.
It influences strongly the time dependence of the fine structure
constant. We have argued that tests of the equivalence principle imply
that the contribution $\sim\beta_{Wn}$ in eq. (\ref{IB}) is small,
such that
\begin{equation}\label{8.1A}
\partial\alpha_{em}/\partial\ln\chi\approx-6.7\cdot10^{-3}\eta_F\end{equation}
If the reported time variation of $\alpha_{em}$ is confirmed,
$\eta_F$ is determined\footnote{We neglect $|B|$ as compared to
$|A|$.} as
\begin{equation}\label{8.2}
\frac{\eta_F}{A}=-3.8\cdot 10^{-6}
\end{equation}
(Otherwise $\eta_F$ should be even smaller in size.)
In a unified theory the variation of the fine structure constant and
the strong gauge coupling are related. Neglecting possible
accidental cancellations in eq. (\ref{GD}), we infer from eq. (\ref{8.2})
an approximate value for
the variation of the ratio between the strong
interaction scale or baryon mass and the Planck mass
\begin{equation}\label{8.3}
|\beta_{ng}|=|\frac{\partial \ln(m_n/\bar M_p)}{\partial\ln\chi}|
\approx1.4\cdot 10^{-4}|A|\end{equation}

The second quantity, $B=\partial\ln f/\partial\ln \chi$, modifies
the ratio between the gravitational mass scales $\bar M_p$
and all particle physics mass scale. In particular, it gives
an additional contribution $\beta_{ng}=-B+...$. Tests of
composition-dependent gravity like forces measure the strength
of the ``fifth force'' mediated by the cosmon. In particular,
the source of the cosmon field of the earth is proportional $\beta_{ng}$.
From present bounds on the differential acceleration between two
test bodies of equal mass but different composition we have concluded
that $|\beta_{ng}|$ cannot be much larger than its value (\ref{8.3})
arising from $\eta_F$. Unless there is a cancellation this gives a similar
bound on $B$, cf. (\ref{W17}). We therefore learn that $B$
must be much smaller in size than our third parameter $A=-\partial\ln V/
\partial\ln\chi+4$.
\begin{equation}\label{8.4}
|B|\stackrel{\scriptstyle<}{\sim}6\cdot10^{-4}k^2|A|\end{equation}
We conclude that the variation in the ratio
between particle masses and $\bar M_p$ is only a small effect for the
cosmological evolution. The dynamics of quintessence depends
only on $A$ and the scalar kinetic term, i.e. $\delta$
or $Z_\chi/f^2$. Actually, only very little is known about the shape of the
cosmon potential or the parameter $A$. It may be as large as
$A=4$ for $V(\chi\to\infty)\to const$, or $A=2$ in case of a small mass term in the
potential, $V(\chi)=m^2\chi^2$ \cite{CWCF}. It could also be in the vicinity
of zero and even be negative. (For $A<0$ the cosmon field $\chi$
decreases towards zero for increasing time.) The cosmological dynamics
of quintessence only depends on the ratio $k^2\approx \delta/A^2\approx 6Z_\chi/(A^2f^2)$.

Finally, we have also investigated the effect of a possible
change in the ratio between the characteristic scales of
weak and strong interactions. In order to keep the discussion
simple, this was done in a very simplified setting where we have
assumed that all particles whose masses arise from the Higgs
mechanism show the same dependence on $\chi$ (i.e. $M_W\sim m_e\sim m_u
\sim m_d$). A rather severe bound (\ref{W19})  arises from tests
of the composition dependence of a fifth force, eq. (\ref{7.14}).
\begin{equation}\label{8.5}
|\beta_{Wn}|\stackrel{\scriptstyle<}{\sim}7\cdot 10^{-6}k^2|A|\end{equation}
Since $\beta_{Wn}$
has a contribution $2\pi\eta_F/(9\alpha_u)=-10^{-4}A$ we conclude
that a substantial cancellation between the two terms in eq. (\ref{GA})
seems necessary unless $k\stackrel{\scriptstyle>}{\sim}3$. This raises
a question of naturalness for quintessence models with $k\stackrel
{\scriptstyle<}{\sim}2$: somehow the physics  determining fermion
masses and the weak scale should ``know'' about the strong scale such
that ratios become only very weakly dependent on $\chi$.

Another question of naturalness concerns the small value of
$\eta_F$ that corresponds to the reported value of $\Delta\alpha
_{em}$. It implies a very weak dependence of the grand unified
gauge coupling on the unification scale. As a comparison,
a ``QCD-like'' $\beta$-function $\beta_g/g\sim g^2/16\pi^2\approx 2\cdot 10^{-3}$
is much larger than the value $\beta_g/g\approx 8\cdot 10^{-6}A$
corresponding to eq. (\ref{8.2}) -- the latter is rather of the
order $(g^2/16\pi^2)^2$. Already the small value of $|\eta_F|$
corresponding to the QSO-observation of $\Delta\alpha_{em}$ calls for an
explanation. For example, the association of $\chi$ with one of
the light scalar fields in a generic string vacuum typically leads to
a much stronger $\chi$-dependence of the gauge coupling. A much smaller
value of $|\eta_F|$ would even be harder to understand! A way out of this
problem may be the attraction of an effective fixed point in the $\chi$-
dependence of $g$ \cite{CWCF}. This could naturally explain a small value
of $|\eta_F|$.

\section{Varying fundamental constants and observations on earth}

Beyond cosmology, additional bounds on the time variation
of coupling constants and mass ratios arise from observations
on earth. For example, the Oklo natural reactor constrains \cite{Oklo}
the variation per year for combinations of the electromagnetic
fine structure constant and mass ratios characteristic for nuclear
cross sections and decays as $m_\pi/m_n$ or $M_W/m_n$.

The issue if the time variation of couplings and mass ratios
on earth should be the same as in cosmology needs some discussion.
This concerns the question if the time variation of the cosmon
field is the same on earth as in the whole universe. We emphasize
that for the metric this is not the case: once a gravitationally
bound object has decoupled from the cosmological evolution, the
time variation of the metric in a local neighborhood around the
object does not reflect the cosmological evolution of the metric
any more. The gravity field near the earth is essentially stationary
-- its time variation corresponds to motions in our solar
system and has nothing to do with the cosmological time evolution.
If the situation would be same for the cosmon field, we would not
expect a time variation of fundamental constants on earth.

We will see, however, that the cosmon field behaves very different
from the metric. Its time variation reflects indeed the cosmological
evolution, even around and within objects like the earth. The origin
of this different behavior arises from the role of the scalar
potential which is absent for the metric. To see this, we start from
the scalar field equation (in the limit of constant $k(\varphi)=k$) in a local
region around a compact object
\begin{equation}\label{9.1}
-k^2D^2\varphi+V'(\varphi)=\frac{\tilde\beta}{\bar M_p}\rho\end{equation}
Here $\rho$ is the density of a spherically symmetric body
(like the earth) which we take for simplicity as $\rho=\rho_E$
for $r<R$ and $\rho=0$ for $r>R$. Correspondingly, the coefficient
$\tilde\beta$ in eq. (\ref{9.1}) reflects the cosmon coupling
to the matter of the compact object. For the earth it is directly
related to the cosmon charge discussed in sect. 7,
\begin{equation}\label{9.2}
\tilde\beta=-\bar M_p\frac{\partial}{\partial\varphi}\ln
\frac{m_n}{\bar M_p}=-\frac{\beta_{ng}}{A+4B}\end{equation}
We note that in general $\tilde\beta\rho_E$ may depend on $\varphi$.

It is convenient to write $\varphi$ in the form
\begin{equation}\label{9.3}
\varphi=\bar\varphi(t)+\varphi_\ell(r,t)+\varphi_\epsilon(r,t)\end{equation}
where $\bar\varphi(t)$ is the homogenous cosmological ``background''
solution obeying
\begin{equation}\label{9.4}
k^2(\ddot{\bar\varphi}+3H\dot{\bar\varphi})+V'(\bar\varphi)=
\frac{\bar\beta}{\bar M_p}\bar\rho(t)\end{equation}
and $\varphi_\ell$ is the local quasistationary cosmon field which is the
analogon to the gravity field of the earth. The size of $\varphi_\epsilon$
will determine the validity of an approximate ``superposition solution''
$\varphi\approx\bar\varphi+\varphi_\ell$.
We will see that $\varphi_\ell/\bar M_p$ and $\varphi_\epsilon/\bar M_p$
are much smaller than one (in contrast to $\bar\varphi/\bar M_p$
which has today a value of $\approx$ 276) and we therefore expand
\begin{equation}\label{9.5}
V'(\varphi)=V'(\bar\varphi)+V''(\bar\varphi)(\varphi_\ell+\varphi_\epsilon)
\end{equation}
The ``local cosmon field'' $\varphi_\ell$ is almost stationary and obeys
the quasistationary field equation
\begin{equation}\label{9.6}
\Delta\varphi_\ell-\mu^2(\bar\varphi)\varphi_\ell+\frac{\tilde\beta(
\bar\varphi)}{k^2\bar M_p}\rho(r)=0\end{equation}
with
\begin{equation}\label{9.7}
\mu^2(\bar\varphi)=\frac{V''(\bar\varphi(t))}{k^2}\end{equation}
The mass term is small as compared to $R^{-2}$, i.e.
$\mu R\approx HR\ll 1$, and we may neglect it. Then the equation
for $\varphi_\ell$ is analogous to Newtonian gravity and one finds
\begin{equation}\label{9.8}
\varphi_\ell(r,t)=\left\{\begin{array}{lll}
\frac{\tilde\beta\rho_E}{2\bar M_pk^2}(R^2-\frac{1}{3}r^2)&{\rm for}&
r<R\\
\frac{\tilde\beta\rho_ER^3}{3\bar M_pk^2r}&{\rm for}& r>R\end{array}
\right.\end{equation}
This analogy between the local cosmon field of the earth, $\varphi_\ell$,
and Newtonian gravity was underlying our discussion of
composition-dependent gravity like forces (or the ``violation
of the equivalence principle'') discussed in sect. 7. In consequence,
the cosmon charges and $\rho_E$ are evaluated at the present value
of $\bar\varphi$. We emphasize, however, that the local cosmon field
alone (i.e. $\varphi\approx\varphi_\ell$) would be inacceptable
as an approximation to the field equation (\ref{9.1}) since
$V'(\varphi_\ell)=-\bar M^3_p\exp(
-\frac{\varphi_\ell}{\bar M_p})$ is many orders of magnitude larger than
$\tilde\beta\rho_E/M_p$. The cancellation of the force term $\sim V^{\prime}$
needs the presence of the background field $\bar{\varphi}$ also in the
gravitationally ``decoupled'' local region. This constitutes
the essential difference to local gravity. We see how the relevance of the
``global clock'' $\bar{\varphi}(t)$ also for local gravitationally
``decoupled regions'' is directly linked to the presence of $V^{\prime}(\varphi)
\neq 0$. Furthermore, only the solution $\varphi\approx\bar\varphi+\varphi_\ell$
approaches the cosmological value $\bar\varphi$ at large distances
from the local object, as it should be.

In order to get a feeling for the size of $\varphi_\ell$, we evaluate
the local cosmon field at the surface of the earth
\begin{eqnarray}\label{9.9}
\frac{\varphi_\ell(R)}{\bar M_p}&=&
\frac{\tilde\beta\rho_ER^2}{3\bar M_p^2k^2}=\frac{\tilde\beta}{k^2}
\frac{M_E}{4\pi\bar M_p^2R}\nonumber\\
&=&-\frac{2\tilde\beta\Phi_E}{k^2}\approx-2\cdot
10^{-13}k^{-2}
\end{eqnarray}
(Here $\Phi_E=-6.9\cdot10^{-10}$ is the Newtonian potential at the
surface of the earth.) Assuming that $\beta_{ng}$ is dominated by
$\eta_F$ as given by the reported cosmological time variation of
$\alpha_{em}$, one has (cf. eq. (\ref{8.2})) $\tilde\beta=-1.4
\cdot10^{-4}$ which leads to the last estimate. Thus
$|\varphi_\ell/\bar M_p|$ is indeed a very small quantity and the
approximation (\ref{9.5}) is justified.

Finally, the size of $\varphi_\epsilon$ determines the accuracy
of the approximate solution $\varphi\approx \bar\varphi+
\varphi_\ell$. Its field equation
\begin{eqnarray}\label{9.10}
&&\ddot\varphi_\epsilon+3H\dot\varphi_\epsilon-\Delta \varphi_\epsilon+\mu^2\varphi
_\epsilon\nonumber\\
&=&-(\ddot\varphi_\ell+3H\dot\varphi_\ell)-\frac{\bar\beta\rho_{miss}}
{k^2\bar M_p}\nonumber\\
&&+\frac{\partial\ln(\tilde\beta\rho)}{\partial\varphi}_{|
\bar\varphi}\frac{\tilde\beta\rho}{k^2\bar M_p}(\varphi_\ell+\varphi
_\epsilon)\end{eqnarray}
has a source term generated by the time dependence of $\varphi_\ell$.
In presence of a possible coupling $\bar\beta$ of quintessence
(i.e.$\bar\varphi$) to the average cosmological energy density
of matter and radiation $\bar\rho$ there is also a contribution due
to the fact that space is empty in a certain region $(r<r_R)$
around the compact object, i.e.$\rho_{miss}=\bar\rho\theta(r_R-r)$.
Finally, the last term reflects the $\varphi$-dependence of $\tilde
\beta$ and the local density $\rho_E$. We note that the
source term is nonvanishing only within a finite region around the
compact object and we are therefore interested in solutions where
$\varphi_\epsilon$ vanishes outside this region for $r\gg r_R$.

The correction $\varphi_\epsilon$ is negligible in the vicinity
of the compact object if the source term on the r.h.s. of eq. (\ref{9.10})
is much smaller than the source term for $\varphi_\ell$ in eq.
(\ref{9.6}), i.e. $\tilde\beta\rho_E/k^2\bar M_p$. We will see that
this is indeed the case. In leading order (neglecting relative
corrections to $\varphi_\ell$ of the order $\mu R$)
one has for $r=R$ (cf. eq. (\ref{9.8}))
\begin{eqnarray}\label{9.11}
&&\frac{\dot\varphi_\ell}{\varphi_\ell}=\partial_t\ln(\tilde\beta\rho_ER^2)
=\frac{\hat\beta\dot{\bar\varphi}}{\bar M_p}\nonumber\\
&&\hat\beta=\frac{1}{A+4B}\left(\frac{\partial\ln
\tilde\beta}{\partial\ln\chi}+\frac{\partial\ln M_E}{
\partial\ln\chi}-\frac{\partial\ln R}{\partial\ln\chi}\right)\end{eqnarray}
The mass of the earth is proportional to the nucleon mass,
$\partial\ln M_E/\partial\ln\chi=\beta_{ng}$, whereas $R$ is
proportional to a typical atomic size. In our approximation $\tilde\beta$
is essentially constant. We see that the time variation of the
local field is negligible as compared to the cosmological time
variation
\begin{equation}\label{9.12}
\frac{\dot\varphi_\ell}{\dot{\bar\varphi}}=\hat\beta\frac{\varphi_\ell}
{\bar M_p}\approx -2\cdot10^{-17}k^{-2}\end{equation}
Also the term $3H\dot\varphi_\ell\sim \hat\beta H^2\varphi_\ell\sim
(\hat\beta/\bar M_p)(\varphi_\ell/M_p)\bar\rho$ is suppressed as
compared to $\tilde\beta(\rho_E/\bar M_p)$ by tiny factors
$(\bar\rho/\rho_E)(\varphi_\ell/\bar M_p)$ and similar for $\ddot
\varphi_\ell$. The other two source terms are also suppressed
by very small factors $\bar\rho/\rho_E$ and $\hat\beta
\varphi_\ell/\bar M_p$, respectively.

We conclude that in the vicinity of a local compact object the
superposition of the cosmological background solution and a
quasistationary local solution  holds to high accuracy $\varphi\approx
\bar\varphi+\varphi_\ell$. Furthermore, for a static local situation
the time change of the cosmon field is completely dominated by the
cosmological background solution. This establishes $\bar\varphi(t)$
as a ``universal clock'' which determines the time variation of fundamental
coupling constants independently of the details of a particular
static local situation. The time variation of couplings measured
on earth must be the same as anywhere else in the universe! Turning
the argument around: any measurement of a time variation of couplings is
directly connected to the cosmological time evolution of quintessence!

Let us next turn to bounds on the time variation of couplings from the
Oklo natural reactor. We have argued previously that the outcome of any experiment or
physical process can only depend on dimensionless couplings or
ratios of masses. For the Oklo reactor, this holds, for example,
for the relative fraction of $Sm$-isotopes from which bounds
on the time variation of $\alpha_{em}$ are derived. In particular,
it cannot depend on the nucleon mass alone and the time variation
cannot involve an isolated contribution $\sim\dot m_n/m_n$.
After all, we could always use a description in a frame where
$m_n$ is constant. In this respect it is important to remember
that the relevant clock for a given process should measure ``seconds''
in units of the appropriate inverse mass scale. For example,
for radioactive decays the time unit is typically given by a weak
decay rate $\sim m^3_nM_W^{-4}$. The appropriate time unit therefore
also depends on the value of
$\bar\varphi$ and on time. A similar remark applies to the relevant
unit of temperature.

The best way of keeping track of these issues is to express
systematically all observables as dimensionless quantities
which only can depend on dimensionless couplings and mass ratios.
(We have always done so in this paper.) Typically, the Oklo
natural reactor will therefore give bounds on the time variation
of combinations of $\alpha_{em}, m_\pi/m_n, m_e/m_n$ and $M_W/m_n$.
Gravitational effects are unimportant such that ratios like
$m_\pi/\bar M_p$ (or only $m_\pi$ for fixed $\bar M_p$)
are not involved\footnote{In this respect our discussion differs
from \cite{CFR}.}. In view of the small value of $|\beta_{Wn}|$
(eq. (\ref{7.14})) we may explore the hypothesis that the time variation
of $\alpha_{em}$ gives a dominant contribution. Under this condition
one may infer a bound \cite{Oklo} $|\Delta\alpha_{em}/\alpha_{em}|
\stackrel{\scriptstyle<}{\sim}10^{-7}$ for $z\approx 0.13$. This bound
can be compared with our discussion of the QSO result in sect. 7
for which we have assumed $\delta\varphi(z_{\rm QSO})/\bar M_p
\approx -2$. Compatibility would require
\begin{equation}\label{9.13}
-\frac{\delta\varphi(z=0.13)}{\bar M_p}\stackrel{\scriptstyle<}{\sim}
0.03\end{equation}
or
\begin{equation}\label{9.14}
\frac{V(z=0.13)-V(0)}{V(0)}\approx0.13\frac{d\ln V}{dz}\stackrel{\scriptstyle<}
{\sim}0.03\end{equation}
Such a small change could only be consistent with a cosmology where
in the present epoch (say $z<1$) the dark energy is dominated by the
potential term and $\varphi$ moves only very little. More
quantitatively, we can express the bound (\ref{9.14}) in terms of
the equation of state of quintessence $w_h$, i.e.
\begin{equation}\label{9.15}
1+w_h=\frac{\dot{\bar\varphi}^2}{V+\dot{\bar\varphi}^2/2}\approx
\frac{1}{3}\frac{d\ln V}{dz}\stackrel{\scriptstyle<}{\sim}0.08\end{equation}
This bound is comparable to bounds from supernovae \cite{SN} and may
be considered as an example how the combination of two different
observations concerning the time variation of $\alpha_{em}$ (QSO and Oklo)
constrains the acceptable models of quintessence. However, in view
of the many uncertainties, in particular concerning
the influence of possible time variations of mass ratios on the
Oklo bound, the bound (\ref{9.15}) should only be interpreted
as an indication that more robust data on the time variation
of couplings will lead to interesting criteria distinguishing
between different proposed models of quintessence.

For a given equation of state a bound
\begin{equation}
\frac{\Delta\alpha_{em}(z=0.13)}{\alpha_{em}}<10^{-7}\kappa
\end{equation}
can be translated into a bound on $\eta_F/A$ using eqs. (\ref{IB}), (\ref{AB7})
\begin{eqnarray}\label{9.15BB}
\left|\frac{\partial\alpha_{em}}{\partial z}\right|&<&
\frac{10^{-7}\alpha_{em}\kappa}{0.13}=5.6\cdot 10^{-9}\kappa\nonumber\\
\left|\frac{\eta_F}{A}\right|&<&\frac{5.6\cdot 10^{-9}\kappa}{6.7\cdot 10^{-3}G(0.13)A}
\approx\frac{3.2\cdot 10^{-7}\kappa}{1+w_h}
\end{eqnarray}
(We have neglected a possible $z$-dependence of $w_h$ in eq. (\ref{AB7}).)

The local cosmon field $\varphi_\ell$ decays $\sim 1/r$ in
outer space around the earth. One may therefore ask if precision
measurements of fundamental couplings by satellite experiments
could see different values as compared to laboratory experiments.
The sensitivity of satellite experiments may be high, especially
if a possible variation of couplings is correlated with a change
of $r$ according to the orbit. The size of such an effect can
be estimated from eq. (\ref{9.9}). As an example, we compute the
change of $\alpha_{em}$ for an orbit at $r=2R$
\begin{equation}\label{9.16}
\frac{\delta\alpha_{em}}{\alpha_{em}}=-\frac{0.9\eta_F}{A}
\frac{\varphi(2R)-\varphi(R)}{\bar M_p}=\frac{0.9\beta_{ng}\eta_F
\Phi_E}{A^2k^2}
\end{equation}
With the QSO-estimate (\ref{8.2}) and assuming that $\beta_{ng}$
is dominated by $\eta_F$ this yields
\begin{equation}\label{9.17}
\frac{\delta\alpha_{em}}{\alpha_{em}}=3.3\cdot10^{-19}k^{-2}\end{equation}
The effect is tiny. Measuring it, for example by using high
precision ``clocks'' (atomic frequencies depend on $\alpha_{em}$)
constitutes a hard experimental challenge. Using orbits exploring the
larger variations  of the cosmon field of the sun enhances the size
of the signal.

\section{Conclusion}

Quintessence relates various interesting possible phenomena:
a time or space dependence of fundamental constants, composition-dependent
gravity like long-range forces and a dynamical dark energy of the universe, as well as
possible modifications of the energy momentum
conservation for matter and radiation in the cosmological equations.
These issues are connected to the coupling of the cosmon field
to matter and radiation, which can in turn be interpreted in terms of
running couplings at the unification scale. Some of the relations
between the variations of fundamental couplings or mass ratios do not
rely on quintessence but rather on the assumption of grand
unification. However, any field-theoretical description of a
cosmological time variation of fundamental constants almost
necessarily involves a scalar \footnote{A logical alternative is a second rank symmetry
tensor.} field which continues to evolve in time
in the recent cosmological history \cite{DZ}. It is then quite natural
(although not compulsory) to associate the potential and kinetic energy
of this scalar field with dark energy and the scalar field with the cosmon
\cite{CW2}. The field description of the time variation of couplings
constitutes the link to tests of the
equivalence principle. It is also needed for an answer to the
question if the time variation of constants on earth should be
the same as for cosmology (see sect. 9).

In principle, the various facets of the cosmon coupling to matter
and radiation constitute a net of predictions in different areas
of observation. Due to the presence of several unknown parameters like
$\eta_F, B, A, k$ and ``$\beta_{Wn}$'', this net is still rather
loose. Under the assumption of the validity of the QSO observation our
present bounds and values for the various parameters are summarized in sect. 8
and in eq. (\ref{9.15BB}). For this we have assumed that $\eta_F,B,A$ and
$\beta_{Wn}$ have been essentially constant for the relevant cosmological
history. Interesting new features may show up if this is not the case.

For the moment, the reported QSO observation of $\Delta
\alpha_{em}$ seems still consistent with the bounds from other
observations. It comes, however, already near those bounds, especially
for the tests of the equivalence principle and for the limits on the time variation
of couplings inferred from the Oklo natural reactor . A further
improvement of laboratory or satellite experiments concerning
the time or space variation of couplings and the validity
of the equivalence principle would be of great value. If the variation
of $\Delta\alpha_{em}$ reported from the QSO is correct, new
discoveries may be around the corner!

We may take an optimistic attitude and imagine that in a not too
distant future several independent observations would measure
effects of the cosmon coupling to matter and radiation.
What would we learn from it about the fundamental interactions?
First, a consistent picture of time varying fundamental constants
would establish the existence of a new
fifth force beyond gravity, electromagnetism, weak and
strong interactions. Second, the time history of the variation
of couplings directly ``measures'' important aspects of the time history of the
cosmological value of the cosmon field. These aspects of the time history
- for example the equation of state $w_h$ -
would then be established completely independently of the
cosmological observation. Comparing it with the time history
needed for a realistic cosmology would provide us with very
strong tests for the cosmological model!

Already now, the reported
QSO observation of $\Delta\alpha_{em}$, if confirmed, places important
restrictions on quintessence. Models with a rather slow time
evolution of the cosmon field in a recent epoch $(z\stackrel{\scriptstyle<}{\sim}1)$
and a more rapid evolution
in the earlier universe $(z\stackrel{\scriptstyle>}{\sim}1)$
are clearly favored. More quantitatively, we have assumed a variation
for the cosmon field $\delta\varphi(z=2)/\bar M_p\approx -2$ for our
numerical estimates. If the magnitude of the variation would be
much smaller, the corresponding value of $\eta_F$ would be
substantially larger, in contradiction with the tests of the
equivalence principle.

Finally, we have seen that the various cosmon couplings can
be related to the dependence of the couplings on the unification
scale in a unified theory. Measuring the cosmon couplings would 
open a completely new window for the exploration of the physics
at the unification scale!

Note added: Based on the present work a model of crossover quintessence has recently
been proposed that is consistent with all observations and constraints on the time
variation of couplings, tests of the equivalence principle and cosmology \cite{CWNEW}.


\begin{thebibliography}{12}

\bibitem{CW2} C. Wetterich, Nucl. Phys. {\bf B302} (1988) 668
\bibitem{PR} P. J. E. Peebles, B. Ratra, Astrophys. Lett. {\bf 325} (1988)
L17
\bibitem{Q} E. Copeland, A. Liddle, D. Wands, Phys. Rev. {\bf D57} (1998)
4686;\\
P. G. Ferreira, M. Joyce, Phys. Rev. {\bf D58} (1998) 1582;\\
R. Caldwell, R. Dave, P. Steinhardt, Phys. Rev. Lett. {\bf 80} (1998)
1582
\bibitem{SF} C. Wetterich, Nucl. Phys. {\bf B302} (1988) 645
\bibitem{CW3} C. Wetterich, Astron. Astrophys. {\bf 30} (1995) 321
\bibitem{DZ} G. Dvali, M. Zaldarriaga, hep-ph/0108217
\bibitem{CK} T. Chiba, K. Kohri, hep-ph/0111086
\bibitem{Obs} J. K. Webb et al., Phys. Rev. Lett. {\bf 87} (2001) 091301
\bibitem{AV} P. Avelino et al., Phys. Rev. {\bf D64} (2001) 103505
\bibitem{SBM} H. Sandwik, J. Barrow, J. Magueijo, Phys. Rev. Lett. {\bf 88} (2002) 031302
\bibitem{OP} K. Olive, M. Popelov, hep-ph/0110377
\bibitem{CFR} X. Calmet, H. Fritzsch, Eur. Phys. J. {\bf C24} (2002) 639,
Phys. Lett. {\bf B540} (2002) 173, hep-ph/0211421
\bibitem{LSS} P. Langacker, G. Segre, M. J. Strassler, hep-ph/0112233
\bibitem{DF} T. Dent, M. Fairbairn, hep-ph/0112279
\bibitem{Uz} J.-P. Uzan, hep-ph/0205340
\bibitem{DV} T. Damour, F. Piazza, G. Veneziano, Phys. Rev. Lett. {\bf 89} (2002) 081601;
Phys. Rev. {\bf D66} (2002) 046007
\bibitem{DMJ} P. M. Dirac, Nature {\bf 192} (1937) 235;\\
E. A. Milne, Proc. Roy. Soc. {\bf A3} (1937) 242;
\bibitem{Jo} P. Jordan, Naturwiss. {\bf 25} (1937) 513; Z. Phys. {\bf 113} (1939) 660
\bibitem{LL} P. Forg\'acs and Z. Horv\'ath, Gen. Relativ. Gravit. {\bf 11}, 205
(1979);\\
W. Marciano, Phys. Rev. Lett. {\bf 52}, 489 (1984);\\
J. D. Barrow, Phys. Rev. D {\bf 35}, 1805 (1987);\\
T. Damour and A. M. Polyakov, Nucl. Phys. {\bf B423}, 532 (1994);\\
Y. Fujii, M. Omote, and T. Nishioka, Prog. Theor. Phys. {\bf 92}, 521 (1994);\\
L.-X. Li and J. R. Gott III, Phys. Rev. D {\bf 58}, 103513 (1998)
\bibitem{HSW} C. T. Hill, Phys. Lett. {\bf B135} (1984) 47;\\
Q. Shafi, C. Wetterich, Phys. Rev. Lett. {\bf 52} (1984) 875
\bibitem{Bek} J. D. Bekenstein, Phys. Rev. {\bf D25}(1982) 1527
\bibitem{EOW} J. Ellis, S. Kalara, K. Olive, C. Wetterich,
Phys. Lett. {\bf B228} (1989) 264
\bibitem{PSW} R. Peccei, J. Sola, C. Wetterich, Phys. Lett.
{\bf B195} (1987) 183
\bibitem{EOT}  S. Baessler et al., Phys. Rev. Lett. {\bf 83} (1999) 3585
\bibitem{Oklo} A. I. Shylakhter, Nature {\bf 264} (1976) 340;\\
T. Damour, F. Dyson, Nucl. Phys. {\bf B480} (1996) 37;\\
Y. Fujii et al., Nucl. Phys. {\bf B573} (2000) 377
\bibitem{CWCF} C. Wetterich, hep-th/0210156
\bibitem{GL}  J. Gasser, H. Leutwyler, Phys. Rep. {\bf 87} (1982) 77
\bibitem{HW} A. Hebecker, C. Wetterich, Phys. Lett. {\bf B497} (2001) 281
\bibitem{NSA} E. Kolb, M. Perry, T. Walker, Phys. Rev. {\bf D33} (1986) 869;\\
L. Bergstr\"om, S. Iguri, H. Rubinstein, Phys. Rev. {\bf D60} (1999) 045005\\
K. Nollet, R. Lopez, Phys. Rev. {\bf D66} (2002) 063507\\
K. Ichikawa, M. Kawasaki, hep-ph/0203006
\bibitem{Mar} J. P. Martins et al., Phys. Rev. {\bf D66} (2002) 023505
\bibitem{Bar} J. Barrow, D. Mota, gr-qc/0212032
\bibitem{SN} P. M. Garnavich et al., ApJ {\bf 509} (1998) 74\\
G. Efstathiou, MNRAS 310(1999) 842
\bibitem{CWBR} C. Wetterich, astro-ph/0111166
\bibitem{CWNEW} C. Wetterich, hep-ph/0301261
\end{thebibliography}
\end{document}